\documentclass[twocolumn,epjc3]{svjour3}  

\journalname{Eur. Phys. J. A}

\usepackage{graphicx}
\usepackage{dcolumn}
\usepackage{bm}
\usepackage{multirow}
\usepackage{amssymb}
\usepackage{amsmath, bbm}
\usepackage[figuresright]{rotating}
\usepackage{capt-of}

\newcommand{\La}{{\Lambda}}
\newcommand{\Si}{{\Sigma}}

\newcommand{\be}{\begin{eqnarray}}
\newcommand{\ee}{\end{eqnarray}}

\newcommand{\nucl}[3]{$^{#1}_{#2}${#3}}

\newlength{\feynwidth} 
\newlength{\feynwidthbig} 

\usepackage{xcolor}

\usepackage{hyperref}
\usepackage{graphicx}

\hypersetup{colorlinks=true,citecolor= blue}

\usepackage[english]{babel}
\usepackage[autolanguage]{numprint}

\usepackage{booktabs}
\usepackage{cite}

\begin{document}



\title{Separation energies of light $\Lambda$ hypernuclei and their theoretical
uncertainties}

\author{Hoai Le\thanksref{addr1,e1}
\and Johann Haidenbauer\thanksref{addr1,e2}
\and Ulf-G. Mei{\ss}ner\thanksref{addr2,addr1,addr3,addr4,e3}
\and Andreas Nogga\thanksref{addr1,addr4,e4}
}
\thankstext{e1}{e-mail: h.le@fz-juelich.de}
\thankstext{e2}{e-mail: j.haidenbauer@fz-juelich.de}
\thankstext{e3}{e-mail: meissner@hiskp.uni-bonn.de}
\thankstext{e4}{e-mail: a.nogga@fz-juelich.de}

\institute{{Institut f\"ur Kernphysik, Institute for Advanced Simulation 
and J\"ulich Center for Hadron Physics, Forschungszentrum J\"ulich, 
D-52425 J\"ulich, Germany} \label{addr1}
           \and
           {Helmholtz-Institut~f\"{u}r~Strahlen-~und~Kernphysik~and~Bethe~Center~for~Theoretical~Physics,
~Universit\"{a}t~Bonn,~D-53115~Bonn,~Germany} \label{addr2}
           \and 
           {Tbilisi State University, 0186 Tbilisi, Georgia} \label{addr3}
           \and 
           {CASA, Forschungszentrum J\"ulich, D-52425 J\"ulich, Germany} \label{addr4}
}

\date{Aug 3, 2023}

\maketitle

\begin{abstract}
Separation energies of light $\Lambda$ hypernuclei ($A\leq 5$)
and their theoretical uncertainties are investigated. Few-body 
calculations are performed within the Faddeev-Yakubovsky scheme and 
the no-core shell model. Thereby, modern and up-to-date nucleon-nucleon, three-nucleon  and  hyperon-nucleon  potentials derived within chiral effective field theory are 
employed. 
It is found that the numerical uncertainties of the few-body methods 
are well under control and an accuracy of around 
$1$~keV for the hypertriton and of better than $20$~keV for the
separation energies of the $^{\,4}_{\Lambda}\mathrm{He}$ and 
$^{\,5}_{\Lambda}\mathrm{He}$ hypernuclei can be achieved. 
Variations caused by differences in the nucleon-nucleon interaction
are in the order of $10$~keV for $^{\,3}_{\Lambda}\mathrm{H}$
and no more than $110$~keV for $A=4,\,5$ $\Lambda$ hypernuclei, when recent high-precision
potentials up to fifth order in the chiral expansion are
employed. The variations are smaller
than the expected contributions from chiral hyperon-nucleon-nucleon  forces  
which arise at the chiral order of state-of-the-art hyperon-nucleon potentials. 
Estimates for those three-body forces  are deduced from a study of the truncation 
uncertainties in the chiral expansion. 
\keywords{Hyperon-nucleon interaction, 
Effective field theory, Hypernuclei}
\end{abstract}

\maketitle

\section{Introduction} 

The insight into the properties of the $\Lambda N$ interaction that 
one can gain from the available scattering experiments 
\cite{Alexander:1969cx,SechiZorn:1969hk,Kadyk:1971tc,Hauptman:1977hr,CLAS:2021gur}
is somewhat limited. Specifically, essential features like its 
spin dependence cannot be deduced from those data. Because of
that, already at an early stage of hypernuclear physics, measurements 
of light $\Lambda$ hypernuclei were explored as an additional source of 
information. For example, 
the conjecture that the $\Lambda N$ interaction in the spin-singlet state
should be more attractive than the one in the triplet state was drawn from
such analyses around 60 years ago \cite{Dalitz:1958zza,DESWART1962458,DIETRICH1964177}.  

Light hypernuclei continue to play an essential role in testing
and improving our understanding of the 
\newline
hyperon-nucleon ($Y\!N$) interaction. Fortunately, the 
theoretical and computational tools for pertinent investigations have
improved dramatically over the years. Techniques for 
treating few-body systems have matured to a level that rigorous 
calculations with sophisticated two-body potentials, including the 
full complexity of  $Y\!N$  dynamics like tensor forces or 
the important coupling between the $\La N$ and $\Si N$ channels, have 
become feasible. 
To be concrete, binding energies of $A=3$ and $4$ hypernuclei can be
obtained by solving Faddeev or Yakubovsky (FY) equations 
\cite{Miyagawa:1993rd,Miyagawa:1995sf,Nogga:2001ef} for such $Y\!N$
interactions.
Other {\it ab initio} methods like the no-core shell model (NCSM) allow one 
to compute even binding energies for hypernuclei beyond the $s$ shell 
\cite{Wirth:2014apa,Wirth:2016iwn,Wirth:2017bpw,Wirth:2018ho,Liebig:2015kwa,Le:2019gjp}
and, so far, studies of hypernuclei up to $^{13}_{\ \La} {\rm C}$ 
have been reported \cite{Wirth:2018ho}. 

With the improvement in the methods another aspect moved into the foreground 
of hypernuclear studies, namely that of estimating the uncertainties of
the achieved results. Of course, this concerns first of all the applied
techniques themselves. However, it extends also to uncertainties due to 
an essential input in such microscopic calculations, the underlying 
nucleon-nucleon ($N\!N$) potential and possibly three-nucleon forces ($3N\!F$s). 
Only with those ingredients under control, reliable conclusions on the $Y\!N$
interaction to be examined can be drawn. 

Bound-state calculations performed over the last two decades suggest that 
the $\La$ separation
energies of light hypernuclei are not very sensitive to the 
employed $N\!N$ interaction \cite{Nogga:2001ef,Haidenbauer:2019boi}.
For example, for a high-precision $N\!N$ interaction derived within 
chiral effective field theory (EFT) like the 
semi-local momentum-space-regula\-rized (SMS) potential of
fifth order (N$^4$LO) \cite{Reinert:2017usi}, 
the variation of the separation energy
with regulator cutoffs $\La_{\mathrm{N}}=400-550$~MeV is of the order of 
$100$~keV for $^{\,4}_{\Lambda}\mathrm{He}/^{\,4}_{\Lambda}\mathrm{H}$ \cite{Haidenbauer:2019boi},
when combined with next-to-leading order (NLO) $Y\!N$ potentials derived 
likewise in chiral EFT. The variation has to be seen in relation to
the total experimental separation energy which is
$2.347 \pm 0.036$~MeV for the $^{\,4}_{\Lambda}\mathrm{He}\, (0^+)$ state
\cite{HypernuclearDataBase}. Indeed, 
such a variation is within the range expected from earlier calculations based 
on phenomenological $N\!N$ and $Y\!N$ interactions \cite{Nogga:2001ef}. 
It was therefore very surprising that Htun et al. \cite{Htun:2021jnu} reported an 
$N\!N$-interaction dependence of the hypertriton separation energy of $100$~keV,
i.e. of the same order as the separation energy itself. 
A recent more extended study by the same group
found variations of around $250$~keV for $A=4$ hypernuclei and of more 
than $1$~MeV for $^{\,5}_{\Lambda}\mathrm{He}$ \cite{Gazda:2022fte}. 
Since the empirical $^{\,5}_{\Lambda}\mathrm{He}$ separation
energy is $3.102 \pm 0.030$~MeV~\cite{HypernuclearDataBase} such a value
implies that the uncertainty could increase dramatically with $A$. 
It should, however, be noted that the analysis in question is limited to $N\!N$ and three-nucleon  ($3N$)
forces of next-to-next-to-leading order (N$^2$LO) in the chiral expansion
and to leading order (LO) with regard to the $Y\!N$ interaction. In the 
earlier work of Wirth and Roth \cite{Wirth:2019cpp} 
variations of 
$\approx 200$~keV and $\approx 400$~keV have been found for $^7_\Lambda$Li and 
$^9_\Lambda$Be, respectively, utilizing N$^3$LO and N$^4$LO $N\!N$ potentials (and N$^2$LO $3N\!F$s), 
but again only LO interactions for the $Y\!N$ system. 

In the present work, we want to re-examine the uncertainties of calculations 
for the separation energies of light hypernuclei. One of 
the main motivations for the study comes from the already mentioned fact that 
the analysis by Gazda et al.~\cite{Gazda:2022fte} is based on two-body potentials of fairly low chiral orders, 
a factor which could limit its conclusiveness. 
As a matter of fact, and as likewise mentioned above, with regard to the 
$N\!N$ system, potentials up to N$^4$LO 
\cite{Reinert:2017usi,Entem:2017hn} are now the standard for 
computations of few-nucleon systems \cite{LENPIC:2022cyu}. 
Also, for the $Y\!N$ interaction, LO is no longer the state-of-art. 
Recently potentials up to N$^2$LO in the chiral expansion have become 
available~\cite{Haidenbauer:2023qhf}. 
Thus, contrary to the calculation in \cite{Gazda:2022fte} where
the focus was on exploring solely a large family of N$^2$LO $N\!N$ and $3N$ potentials, 
called NNLO$_{sim}$ \cite{Carlsson:2015vda}, 
we extend our analysis to variations observed when employing potentials 
of different chiral order, for the $N\!N$ as well as the $Y\!N$ systems. 
Specifically, we consider $N\!N$ potentials from LO to 
N$^4$LO, supplemented by $3N\!F$s starting from N$^2$LO,  
and $Y\!N$ potentials from LO to N$^2$LO. 

The paper is structured in the following way: In the subsequent
section we describe the strategy of our analysis and
the $N\!N$, $3N$, and $Y\!N$ interactions used as input. 
In Sect.~\ref{Sec:Method} uncertainties related to the applied methods 
for treating few-body systems are explored. Uncertainties due to the
employed $N\!N$ and $Y\!N$ interactions are investigated in 
Sect.~\ref{Sec:Potentials}. The paper closes with a brief
summary. 

\section{Strategy and input}
\label{Sec:Strategy}

In the uncertainty analysis for $N\!N$ interactions derived within chiral
EFT, several aspects have been considered such as the error due to the 
truncation in the 
chiral expansion, statistical uncertainties in the LECs of the $N\!N$
contact terms, errors associated with the pion-nucleon LECs, 
and the role of the energy range when fitting the $N\!N$
scattering data 
\cite{Epelbaum:2014efa,Furnstahl:2014xsa,Furnstahl:2015rha,Carlsson:2015vda,Reinert:2017usi}. 
In general the uncertainty is dominated by the truncation error. 
Thus, in our analysis for the $Y\!N$ interaction the main focus will be likewise 
on the uncertainty due to the truncation in the chiral expansion. 
We employ the most advanced chiral $N\!N$ potentials (N$^4$LO$^+$)
of the Bochum group, 
which provide the presently best possible representation of the $N\!N$ 
interaction, for the main part of our analysis. Here
the $^+$ in N$^4$LO$^+$ indicates that some of the short-range
operators appearing at N$^5$LO are also included, see \cite{Reinert:2017usi}.

Our study is to some extend complementary to the work of
Refs.~\cite{Htun:2021jnu,Gazda:2022fte} where the focus was
on a statistical exploration of effects from the nuclear interactions, 
based on a family of 42 $N\!N$ and $3N$ N$^2$LO potentials.
We restrict the number of variations and combinations of 
$N\!N$ $(3N)$ and $Y\!N$ potentials, in view of our limited CPU resources, and give priority to precision and reliability of 
the computation within the FY  and the Jacobi-NCSM (J-NCSM) methods \cite{Le:2020zdu}. 
Accordingly, we perform only selective calculations with 
lower-order $N\!N$ potentials for orientation and illustration - and also to 
connect with some of the results presented in  Refs.~\cite{Htun:2021jnu,Gazda:2022fte}. 
Of course, strict compliance with the power counting would require that we
treat the $N\!N$ and $Y\!N$ systems on the same level in studies of hypernuclei, 
i.e. combine a LO $Y\!N$ potential with a LO $N\!N$ potential, etc. 
However, we think that this procedure would provide little insight into the properties of the $Y\!N$ force, given the fact that we are not able to achieve the same accuracy for $Y\!N$ 
as for the $N\!N$ interaction. 
Thus, we believe that the strategy that we follow here minimizes the bias
from the $N\!N$ potential and allows for the best possible estimate of 
the truncation error for the hypernuclear separation energies due 
to the $Y\!N$ interaction. 

\begin{table*}[tbp]
    \centering
    \setlength{\tabcolsep}{2ex}
    \renewcommand{\arraystretch}{1.5}
    \begin{tabular}{|l|r|rrr|rr|}
    \hline
        order  & $\Lambda_N$ & $c_1$  & $c_3$  & $c_4$  & 
         $c_D$  & $c_E$ \\
         \hline \hline
        N$^2$LO   &  400   & -0.74 & -3.61 & 2.44 &  8.0069 & -0.94276  \\
        N$^2$LO   &  450   & -0.74 & -3.61 & 2.44 &  2.4850 & -0.52793  \\
        N$^2$LO   &  500   & -0.74 & -3.61 & 2.44 & -1.6262 & -0.062696 \\
        N$^2$LO   &  550   & -0.74 & -3.61 & 2.44 & -6.6840 &  0.85320  \\
        \hline 
        N$^3$LO   &  400   & -1.20 & -4.43 & 2.67 &  3.9998 & -0.45796 \\
        N$^3$LO   &  450   & -1.20 & -4.43 & 2.67 &  1.5281 & -0.35397 \\
        N$^3$LO   &  500   & -1.20 & -4.43 & 2.67 & -0.4344 & -0.31055 \\
        N$^3$LO   &  550   & -1.20 & -4.43 & 2.67 & -2.7685 & -0.12613 \\
        \hline 
        N$^4$LO   &  400   & -1.23 & -4.65 & 3.28 &  3.1275  & -0.44217 \\
        N$^4$LO   &  450   & -1.23 & -4.65 & 3.28 &  0.65260 & -0.35275 \\
        N$^4$LO   &  500   & -1.23 & -4.65 & 3.28 & -1.5794  & -0.32025 \\
        N$^4$LO   &  550   & -1.23 & -4.65 & 3.28 & -3.9867  & -0.32235 \\
        \hline 
        N$^4$LO$^+$   &  400 & -1.23 & -4.65 & 3.28 &  3.3278 &  -0.45405 \\
        N$^4$LO$^+$   &  450 & -1.23 & -4.65 & 3.28 &  0.8918 &  -0.38595 \\
        N$^4$LO$^+$   &  500 & -1.23 & -4.65 & 3.28 & -1.2788 &  -0.38214 \\
        N$^4$LO$^+$   &  550 & -1.23 & -4.65 & 3.28 & -3.6257 &  -0.41022 \\
        \hline
    \end{tabular}
    \caption{Parameters $c_i$, $c_D$ and $c_E$ of the $3N$ interaction adjusted in conjunction with different orders and cutoffs $\Lambda_N$ of the $N\!N$ interaction (see \cite{Maris:2020qne} for details). $\Lambda_N$ is given in MeV, $c_i$ in [GeV$^{-1}$] and $c_D$/$c_E$ are dimensionless. }
    \label{tab:3nlecs}
\end{table*}

In the following, we evaluate the $^{\,3}_{\Lambda}\mathrm{H}$, 
$^{\,4}_{\Lambda}\mathrm{He}$ and $^{\,5}_{\Lambda}\mathrm{He}$
separation energies using the SMS $N\!N$ potential at order $\mathrm{N^4LO}^+$ and 
$3N\!F$s at order $\mathrm{N^2LO}$ (for all the available cutoffs of 
$\Lambda_{N}= 400,\, 450,\, 500,\, 550$~MeV), in combination with the 
SMS $Y\!N$ potentials at orders NLO and N$^2$LO with cutoff $\Lambda_Y=550$~MeV. In addition, for convergence study, we also perform calculations using  the $N\!N$ interactions at lower orders for the cutoff $\Lambda_{N}=450$ MeV and the $Y\!N$ interaction at LO. 
Information on the employed SMS $N\!N$ potentials can be found in
Ref.~\cite{Reinert:2017usi}, while the 
SMS $Y\!N$ potentials are described in Ref.~\cite{Haidenbauer:2023qhf}. The $3N\!F$s are identical to the ones used in \cite{Maris:2020qne,Maris:2023esu}. 
For completeness, we summarize the parameters $c_i$, $c_D$ and $c_E$ in Table~\ref{tab:3nlecs} (see Eq.~(1) in \cite{Maris:2020qne} for the definition). The $c_i$ values used in \cite{Reinert:2017usi} were obtained in Ref.~\cite{Hoferichter:2015tha} from matching the results of a Roy-Steiner analysis of pion-nucleon scattering to chiral perturbation theory.   For the $3N\!F$s at  N$^{3,4}$LO, these values need to be shifted as outlined in \cite{Bernard:2007sp} which is already taken into account in Table~\ref{tab:3nlecs}. The $c_d$ and $c_e$ LECs are fitted to the $^3\mathrm{H}$ binding energy and the  proton-deuteron  differential cross-section minimum at the beam energy of $E_{N} = 70$~MeV \cite{Maris:2020qne}.

Since we will compare to some results from Gazda et al.~\cite{Gazda:2022fte},
the construction and the properties of the potential set 
NNLO$_{sim}$ used in \cite{Gazda:2022fte} will be of relevance for the discussion below. 
The interactions are described in Ref.~\cite{Carlsson:2015vda}. There, six different fitting regions for 
$N\!N$ (namely $T_{lab} = 125, 158, 191, 224, 257, 290$~MeV,
and seven cutoffs $\Lambda_N=450,\, 475, ...,\, 600$~MeV are considered.  
 The LECs of these interactions are optimized by requiring a simultaneous
description of $N\!N$ as well as $\pi N$ scattering cross sections,
and binding energies and charge radii of the deuteron, $^3$H, and $^3$He.
The combined analysis lead to $\pi N$ LECs that are marginally consistent with the ones
from the Roy-Steiner analysis and with much larger uncertainties \cite{Hoferichter:2015tha}. Thus, it is
preferable to use such knowledge
directly as done e.g. in Ref.~\cite{Reinert:2017usi}.

The strategy followed in the construction of the 
SMS $N\!N$ potentials~\cite{Reinert:2017usi} that are employed in our investigation
is however different, cf. Sects.~6.3 and 7.5.4 of that reference for a detailed
discussion. Here a smaller (larger)
energy range was considered for establishing the lower (higher) order 
$N\!N$ potentials, which is in line with the expected pertinent validity range of the 
chiral expansion. Specifically, at N$^2$LO,  N$^3$LO and
N$^4$LO$^+$  
$N\!N$ data up to 125, 200 and 260~MeV, respectively, were fitted.  
The $\chi^2$ obtained in the fit for different orders
and different energy regions are listed in Table~3 of \cite{Reinert:2017usi}
for the cutoff $\Lambda_N = 450$~MeV. With that cutoff the overall best
description of the $N\!N$ data is achieved. Corresponding results for all 
cutoffs can be found in \cite{Reinert:2022jpu} (Table 6.9). One can see
that for the N$^4$LO$^+$ potentials the $\chi^2$/datum is excellent
(close to $1$) and practically  independent of the cutoff, while for 
N$^2$LO 
the quality 
is rather different for the different energy regions and depends strongly
on the cutoff.
Clearly, the $N\!N$ data at high energy cannot be well described by the $N\!N$ interactions at low order.
To the best of our knowledge, there is no detailed information on the $\chi^2$
for the individual potentials of the set NNLO$_{sim}$. 
Finally, note that the NNLO$_{sim}$ potentials are based on a
nonlocal regulator throughout
while in the SMS $N\!N$ potentials a local regulator is employed 
for the pion-exchange contributions. Here the range of optimal cutoffs is $400-550$~MeV \cite{Reinert:2017usi}, as already indicated above. 

\section{Uncertainties from the method}
\label{Sec:Method} 

In the NCSM calculations,
all potentials are evolved by a similarity renormalization group (SRG)
transformation based on a flow parameter of $\lambda=1.88$~fm\textsuperscript{-1}
unless stated otherwise, see 
Refs.~\cite{Le:2020zdu,Le:2022ikc} for the technical details. 
SRG-induced $Y\!N\!N$ forces are taken into account
(and, of course, both SRG-induced and chiral $3N$ forces) but no chiral $Y\!N\!N$ three-body forces ($3B\!F$s). 
In this section, 
 we will discuss the uncertainties of our numerical approaches and  quantify the effect of neglecting the SRG-induced four- and higher-body forces on the $\Lambda$ separation energies $B_\Lambda$ by carefully studying the dependence of the
 $B_{\Lambda}$'s on the SRG-flow parameters. Note that results for the   $A=3(4)$ hypernuclei have been mainly obtained by solving a set of the FY equations, which are formed by rewriting the  corresponding non-relativistic  momentum-space Schr\"{o}dinger equations,  with the bare $N\!N$, $3N$ and $Y\!N$ interactions.  The FY method is more  efficient for light systems and has been very successfully applied to study both nuclei and hypernuclei up to four baryons \cite{Miyagawa:1993rd,Miyagawa:1995sf,Nuclearandhypernuc:2001wd,Nogga:2001ef}.  In addition, it has also been carefully checked that the FY equations converge within less than 1(20) keV for $A=3 (4)$ hypernuclei, respectively (see also \cite{Nogga:2001ef, Haidenbauer:2019boi}). Hence, a direct comparison between the J-NCSM and the FY results for $^{\,4}_{\Lambda}\mathrm{He}$ will  provide the most accurate estimate for the size of the neglected contributions from  the induced higher-body forces in this system. Finally, the FY approach,  in principle,  could also be extended to $^{\,5}_{\Lambda}\mathrm{He}$, however, the computation is rather challenging, therefore, for that system, we will only employ the J-NCSM.  For details of the method and its implementation, we refer the reader to \cite{Liebig:2015kwa,Le:2020zdu,Le:2019gjp}.

       \begin{figure*}[tbp] 
      \begin{center}
     { \includegraphics[width=0.4\textwidth,trim={0.0cm 0.00cm 0.0cm 0 cm},clip]{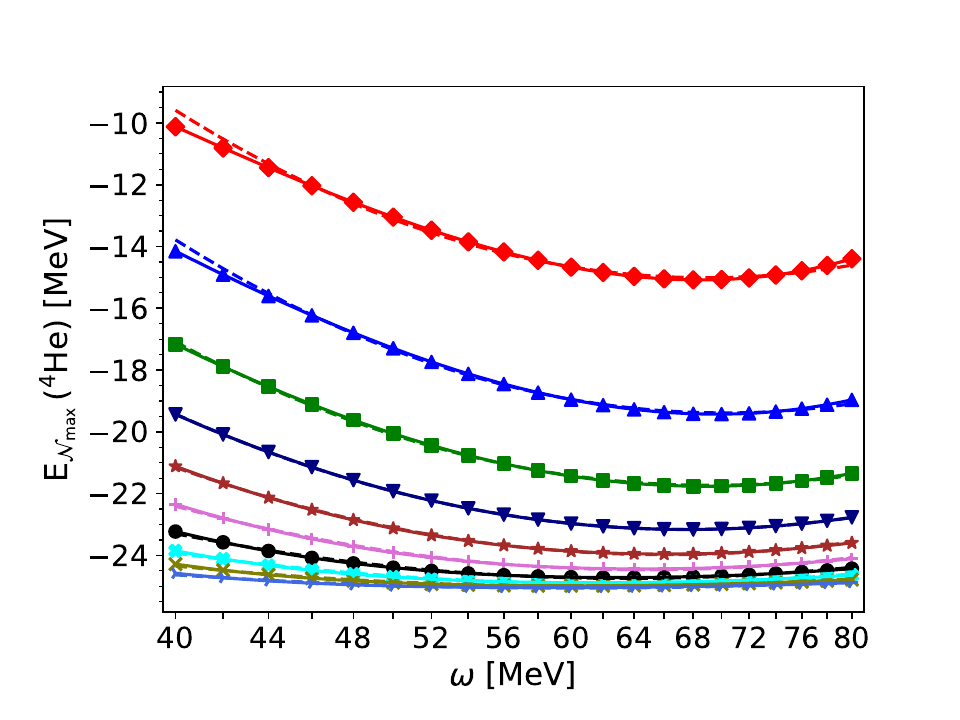}
       \includegraphics[width=0.4\textwidth,trim={0.0cm 0.00cm 0.0cm 0 cm},clip]{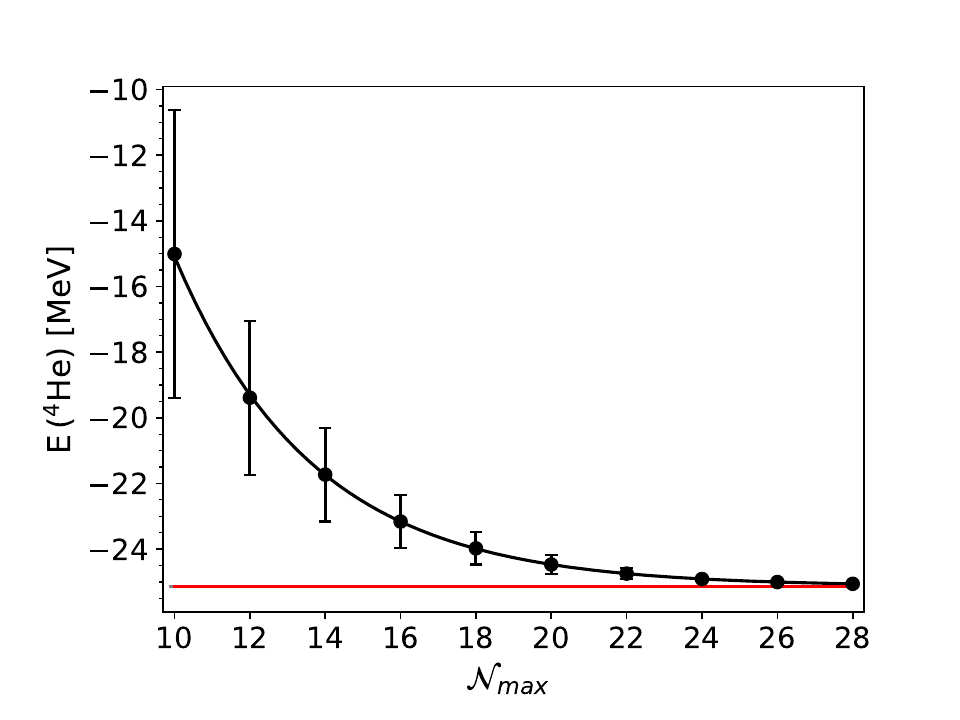}}
        { \includegraphics[width=0.4\textwidth,trim={0.0cm 0.00cm 0.0cm 0 cm},clip]{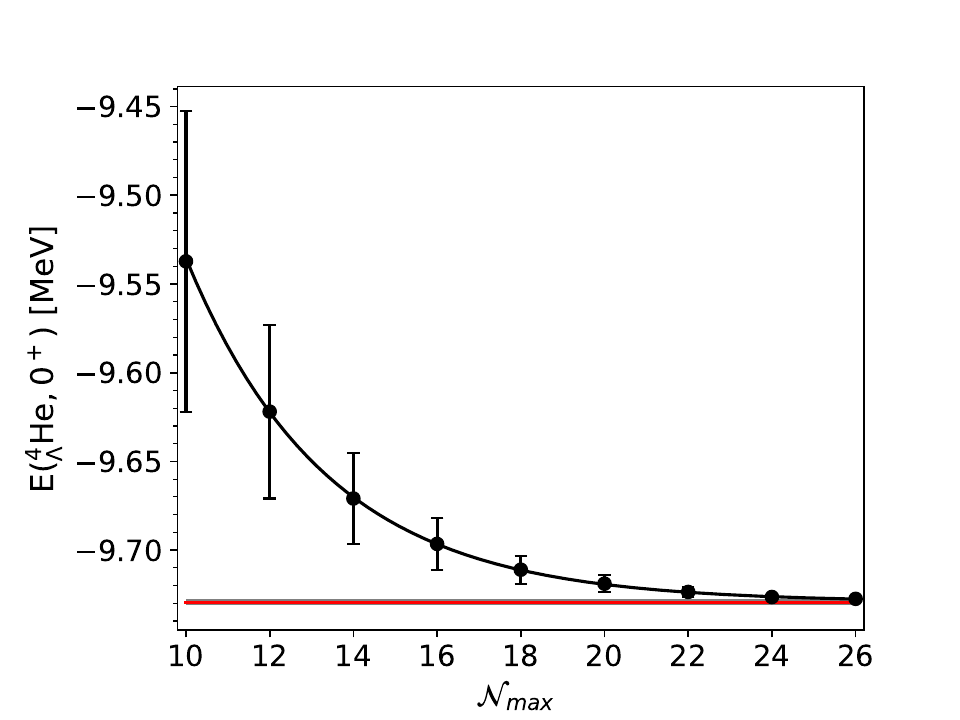}
       \includegraphics[width=0.4\textwidth,trim={0.0cm 0.00cm 0.0cm 0 cm},clip]{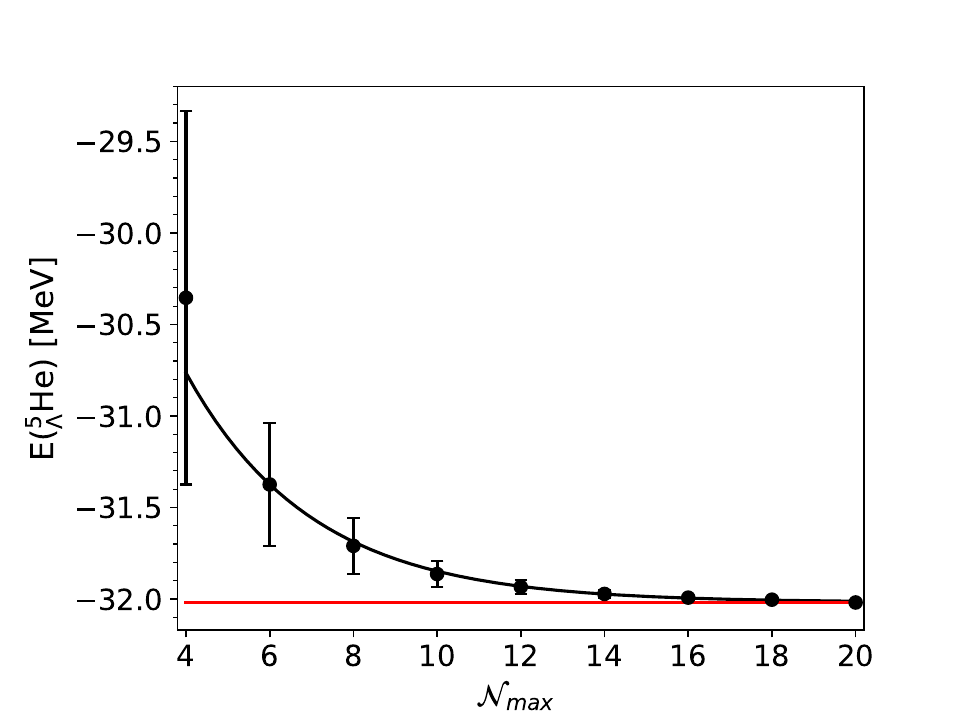}}
      \end{center}
                 \caption{Upper panel: two-step extrapolation procedure for $E({\mathrm{^4He}})$. $\omega$-space extrapolation (left). 
                 The solid lines are the $^4$He binding energies computed for different $\mathcal{N}_{\rm max}$ from 10 to 28 with a step of 2. 
                 The dashed lines are obtained by using the ansatz Eq.~(\ref{eq:ansatz1}).  $\mathcal{N}_{\rm max}$-space extrapolation (right). The horizontal line with shaded area shows the extrapolated binding energy and the estimated numerical uncertainty. The calculations are based on the SMS $\mathrm{N^2LO}(550)$ $N\!N$ potential.  Lower panel: $\mathcal{N}$-space extrapolation for  $E({\mathrm{^4_{\Lambda}He}})$ (left) and $E({\mathrm{^5_{\Lambda}He}})$ (right).  The calculations are based on the SMS $\mathrm{N^4LO^+(450)}$ $N\!N$ potential with $\mathrm{N^2LO(450)}$ $3N$ force, 
                 and the $\mathrm{N^2LO}(550)$ $Y\!N$ potential.}
    \label{fig:extrapolate-E4He}
         \end{figure*}   
\subsection{$\omega$- and $\mathcal{N}_{\rm max}$-space extrapolation}
As  mentioned earlier, we will employ the J-NCSM to calculate the binding energies of  systems with $A \ge  4$. This method is of course  also applicable to $^{\,3}_{\Lambda}$H, however, because of its extremely small  separation energy, the J-NCSM calculations for that system inhere an uncertainty that is significantly larger than the value of 1 keV for the Faddeev method.
  In general, the J-NCSM approach relies   on an expansion of the many-body wavefunction in harmonic oscillator (HO)  basis depending on relative Jacobi coordinates of all the particles involved. Such basis functions are characterized by the HO frequency $\omega$ and the total HO energy quantum number $\mathcal{N}$. In order to get a finite number of basis states for  practical calculations, $\mathcal{N}$ is constrained by the model space size  $\mathcal{N}_{\rm max}$ \cite{Liebig:2015kwa,Le:2020zdu}.  In practice, we perform the J-NCSM calculations for different sets  of all the accessible  model spaces up to $\mathcal{N}_{\rm max}$ and
  for a certain range of HO frequencies (which are close to the variational minimum at $\mathcal{N}_{\rm max}$). The converged binding energies are then obtained by performing an additional extrapolation  to infinite model space. Several strategies haven been pursued to perform such extrapolations. Very often, an empirical exponential extrapolation in $\mathcal{N}_{\rm max}$ at a fixed HO frequency $\omega$  (usually the ones  that yield the lowest binding energy
for the  largest computationally accessible model spaces) is employed \cite{Maris:2023esu,Maris:2008ax,Jurgenson:2013yya,Coon:2012ab}. In our works in \cite{Liebig:2015kwa,Le:2020zdu,Le:2019gjp} and also for this work, we pursue a slightly different strategy, namely a two-step extrapolation procedure.  Here the first step is to minimize (eliminate) the HO-$\omega$ dependence of the binding energies $E(\omega,\mathcal{N}_{\rm max})$ utilizing the following (empirical) ansatz,
\begin{align}\label{eq:ansatz1}
E(\omega,\mathcal{N}_{\rm max}) = E_{\mathcal{N}_{\rm max}}  + \kappa (\log(\omega) - \log(\omega_{opt}))^2,
\end{align}
with $ E_{\mathcal{N}_{\rm max}}, \,\omega_{opt}$ and $\kappa$ being fitting parameters. The obtained lowest energies $E_{\mathcal{N}_{\rm max}}$ for each accessible  $\mathcal{N}_{\rm max}$ are then used  for the extrapolation to infinite model space   assuming  an exponential ansatz,
\begin{align}
E_{\mathcal{N}_{\rm max}} = E_{\infty}  + A  e^{-B\mathcal{N}_{\rm max}}.
\end{align}

The final uncertainty is assigned as the difference between the  infinite-model space extrapolated  energy $E_{\infty}$  and the one computed for the  largest computationally accessible model space. The described two-step extrapolation procedure is applied to all nuclear and hypernuclear 
calculations of the present work.  As for demonstration, we show in  Fig.~\ref{fig:extrapolate-E4He} the  extrapolation  for the $^4\mathrm{He}$  binding energies that have been computed using the SMS $\mathrm{N^2LO(550)}$ NN potential.  Here,  we obtain  a binding energy of $E(^4\mathrm{He})= -25.14 \pm 0.06$~MeV  which is in a good agreement with the value of $E(^4\mathrm{He})= -25.15 \pm 0.02$~MeV that resulted from solving the FY equations.  Clearly, our  way of assigning the numerical uncertainty seems to  be rather  conservative,  and a somewhat less conservative (but also empirical) estimate has been considered for example  in \cite{Maris:2023esu,Maris:2008ax,Jurgenson:2013yya,Coon:2012ab}. 
Nevertheless, as one will see in the following section, using the SRG-evolved interactions, our NCSM results for $A=4,5$ hypernuclei converge almost perfectly (within several keV).

\subsection{ Infrared (IR) extrapolation }
\begin{figure}[tbp] 
      \begin{center}
        { \includegraphics[width=0.4\textwidth,trim={0.0cm 0.00cm 0.0cm 0 cm},clip]{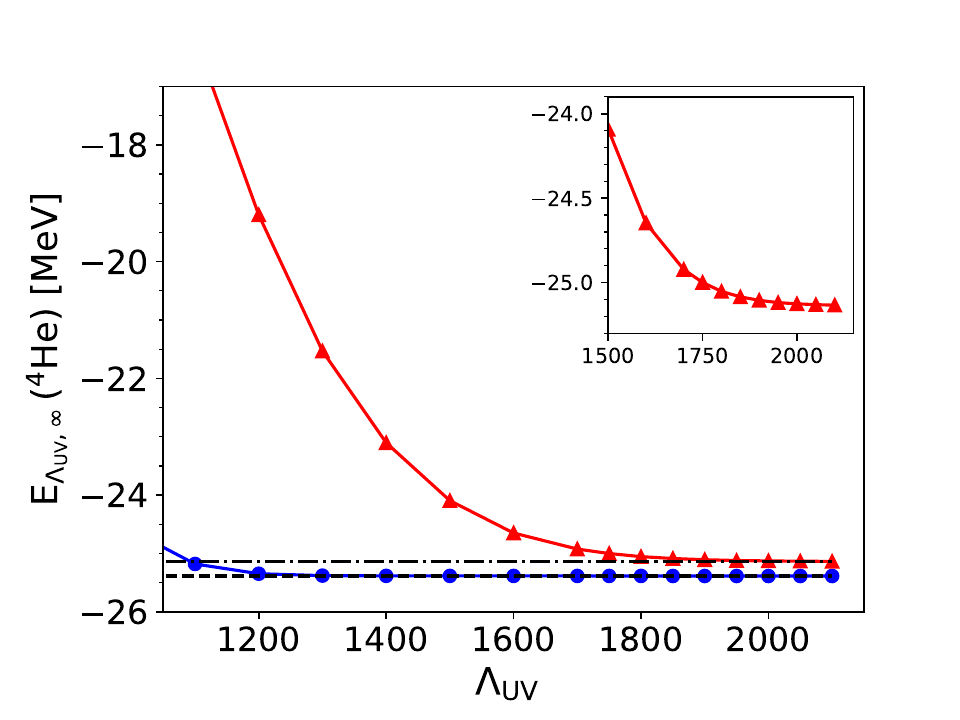}}
      \end{center}
                 \caption{ IR extrapolation based on Eq.~(\ref{eq:IRextra}) of $E(^4\mathrm{He})$ at different  UV cutoffs $\Lambda_{UV}$. The calculations are based on the SMS $\mathrm{N^2LO}(550)$ (red triangles) and Idaho-$\mathrm{N^3LO}(500)$ (blue circles) $N\!N$ potentials. Dash-dotted and dashed  lines are the corresponding binding energies obtained using the extrapolation formula in Eq.~(\ref{eq:ansatz1}). }
    \label{fig:extrapolate-IR-4He}
         \end{figure}

Let us further note that a truncation in the $\omega$ and $\mathcal{N}_{\rm max}$ model
spaces also implies a finite infrared (IR) length scale, ($L_{IR}$), and  an ultraviolet  (UV) cutoff,  $\Lambda_{UV}$,   \cite{Stetcu:2006ey,Jurgenson:2011ec,Coon:2012ab,PhysRevC.90.064007}. Hence, by recasting  the binding energies $E(\omega, \mathcal{N}_{\rm max})$ in terms of $L_{IR}$ and $\Lambda_{UV}$,
$E(L_{IR}, \Lambda_{UV})$, one can also perform the infinite basis extrapolation with respect to the $L_{IR}$ and $\Lambda_{UV}$ cutoffs \cite{Stetcu:2006ey,Jurgenson:2011ec,Coon:2012ab,PhysRevC.90.064007}.  In general, the IR length scale (and $\Lambda_{UV}$)
depends on the system considered  and on how the basis functions are truncated. In the case of the NCSM  with a total energy truncation a   precise value for $L_{IR}$ has been derived in  \cite{Wendt:2015nba}. Furthermore, it has also been shown  that, at a sufficiently large and  fixed  $\Lambda_{UV}$, the leading order IR correction to the binding
energy  follows an exponential dependence on $L_{IR}$ \cite{Coon:2012ab,Furnstahl:2013vda,Wendt:2015nba},
\begin{align}\label{eq:IRextra}
E_{\Lambda_{UV}}^{}(L_{IR}) = E_{\Lambda_{UV}, \infty}^{}  + a_{\Lambda_{UV}}^{}  e^{-2\kappa_{\Lambda_{UV},\infty}^{}L_{IR}} . 
\end{align}
The UV correction is in general sensitive  to the  details of the employed interaction or, more precisely, on how the interaction is regularized  \cite{PhysRevC.90.064007}.  This correction is not yet well understood in contrast to the IR energy correction. Hence, in practice, one often performs the IR extrapolation at a sufficiently large and  fixed UV cutoff which yields reliable IR extrapolations and for which the UV error is approximately minimized (or suppressed) \cite{Forssen:2017wei,Gazda:2022fte}. As an example, we show in Fig.~\ref{fig:extrapolate-IR-4He} the IR extrapolated binding energy of $^4\mathrm{He}$, $E_{\Lambda_{UV},\infty}(^4\mathrm{He})$,  as a function of $\Lambda_{UV}$. The red triangles and blue circles are the binding energies computed using the SMS   $\mathrm{N^2LO} (550)$ and  Idaho-$\mathrm{N^3LO}(500)$ $N\!N$ potentials, respectively. It clearly sticks out that, with the  Idaho-$\mathrm{N^3LO}(500)$ interaction (i.e. the one with a non-local regulator), the IR extrapolated results are practically stable for a sufficiently large  UV cutoff  ($\Lambda_{UV} \ge 1300$ MeV). Indeed, the overall variation of  $E_{\Lambda_{UV}, \infty}(^4\mathrm{He})$ for a range of UV cutoffs of $ 1300 \le \Lambda_{UV} \leq 2100$ MeV is about 1 keV only. In contrast, for the $\mathrm{N^2LO} (550)$ potential with a semi-local regulator, we observed a variation of about 90 keV even  for very large $\Lambda_{UV}$ but in a significantly smaller range, namely   $1800 \leq  \Lambda_{UV} \leq 2100$~MeV (see also the insert plot in Fig.~\ref{fig:extrapolate-IR-4He}). Evidently, the UV correction for the SMS interactions seems to be sizable and therefore should  be carefully studied 
when  the IR extrapolation is being used. 
         
Finally, we have  also adopted a Bayesian approach for the IR extrapolation as recently employed by Gazda et al.~\cite{Gazda:2022fte}. Here, we observed that the extrapolated results are rather sensitive to the hyperparameter chosen for $\Delta E_{\mathrm{IR, max}}$, see Eqs.~(28) an in \cite{Gazda:2022fte}. By choosing $\Delta E_{\mathrm{IR, max}}$ to be twice of the maximum extrapolation distance, we obtained 
the  $^4\mathrm{He}$ binding energies of \newline
$E(^4\mathrm{He})= -25.06 \pm 0.04$  and  $-25.12 \pm 0.04$ MeV for the $\mathrm{N^2LO}(550)$   interaction at very large UV cutoffs of $\Lambda_{UV} =1800$ and 2000 MeV, respectively. One sees that the latter energy  is in a good  agreement with the value of $E(^4\mathrm{He})= -25.14 \pm 0.06$ MeV that
resulted from the two-step $\omega$- and $\mathcal{N}_{\rm max}$ extrapolation
and with the binding energy  of  $E(^4\mathrm{He})= -25.15 \pm 0.02$   obtained by solving the FY equations. Still, there is a non-negligible discrepancy of 60~keV between the two IR extrapolated results at $\Lambda_{UV}= 1800$ and $2000$~MeV which could be attributed to the UV truncation or the high-order IR corrections.  As discussed above, the latter is sensitive to the underlying interactions and their regulator and seems to be particularly significant for the SMS interactions. With the SRG-evolved potentials, the dependence on the chosen UV cutoff is somewhat reduced
but it remains visible.  
In the following, we will therefore  employ the two-step extrapolation to extract the final binding energies for $A=4, 5$ systems. This extrapolation procedure is robust and it  depends neither on the systems investigated nor on the underlying interactions. Let us finally  stress that due to   the SRG evolution and  the very large model spaces employed in the calculations, our computed energies  for $A=4, 5$ hypernuclei at the largest model spaces  practically converge. The final results should therefore not depend on the extrapolations.


 \begin{table*}[tbp]
 
\renewcommand{\arraystretch}{1.5}
\begin{center}
  \setlength{\tabcolsep}{0.6cm}
\begin{tabular}{|l|  r  r|  r|}
\hline
  $\lambda$  [fm\textsuperscript{-1}]   &    \multicolumn{2}{c|}{ $B_{\Lambda}(^4_{\Lambda}\mathrm{He},0^+)$}  &   $B_{\Lambda}(^5_{\Lambda}\mathrm{He})$ \\
  & $\mathrm{N^2LO(550)}$ & NLO(550)      &$\mathrm{N^2LO(550)}$ \\
 \hline
\hline
  1.88    &  1.992  $\pm$ 0.002  & 2.061  $\pm$ 0.001  &  3.712    $\pm$ 0.001 \\
  2.00    &  1.991   $\pm$ 0.005 & \multicolumn{1}{c|}{---} & 3.70$\pm$  0.005\\
  2.236    & 1.990   $\pm$ 0.007 & 2.06   $\pm$ 0.006  &  3.708   $\pm$  0.006\\
  2.60    &  1.989   $\pm$ 0.014  &  2.06   $\pm$ 0.012 & 3.744     $\pm$  0.008\\
  3.00    &  1.985   $\pm$ 0.024 &  2.058   $\pm$ 0.023  & 3.806     $\pm$  0.028\\
  4.00   & \multicolumn{1}{c}{---} &   2.052   $\pm$ 0.021  &   \multicolumn{1}{c|}{---}  \\
 $\infty$    &  2.01 $\pm $ 0.02   &  2.08 $\pm $ 0.02  & \multicolumn{1}{c|}{---}\\
\hline
  \end{tabular}
\end{center}
\renewcommand{\arraystretch}{1.4}
\caption{\label{tab:Blambda_4_5} $\Lambda$ separation energies  $B_{\Lambda}(^4_{\Lambda}\mathrm{He},0^+)$ and $B_{\Lambda}(^5_{\Lambda}\mathrm{He})$ in MeV computed for different SRG flow parameters $\lambda$. All calculations are based on the SMS $N\!N$ potential $\mathrm{N^4LO}^+(450)$ and the
 $3N$ force at $\mathrm{N^2LO}(450)$. The SRG-induced $3N$ and $Y\!N\!N$ forces are also included.  $B_{\Lambda}(^4_{\Lambda}\mathrm{He},0^+)$  at $\lambda=\infty$ is obtained by solving the FY equations employing the bare $N\!N$, $3N$ and $Y\!N$ potentials. Note that $B_{\Lambda}(^4_{\Lambda}\mathrm{He},0^+)$  at $\lambda= 4.0 \,(3.0)$ fm\textsuperscript{-1} have been computed for model spaces up to $\mathcal{N}_{\rm max} =34 \,(28)$, respectively, whereas values at lower $\lambda$ are computed for  $\mathcal{N}_{\rm max} =26$.}
\end{table*}

\subsection{Similarity Renormalization Group (SRG) for $^{\,4}_{\Lambda}\mathrm{He}$, $^{\,5}_{\Lambda}\mathrm{He}$}
In order to study the different extrapolation methods, we have  employed so far only the bare two-body $N\!N$ interactions and provided examples for the $^4\mathrm{He}$ system. There, one can clearly see that the NCSM calculations converge nicely even when the bare chiral $N\!N$ interaction is employed.  Note that  $3N$ forces have been omitted in such calculations in order to save computational resources, but we do not expect that these change the convergence of the NCSM calculations significantly.
However, when a hyperon is added to the $A=3\, (4)$ nuclear systems, the NCSM calculations for $^{\,4}_{\Lambda}\mathrm{He}$ ($^{\,5}_{\Lambda}\mathrm{He}$) with the bare  chiral  SMS $N\!N$, $3N$ and $Y\!N$ interactions do not converge well even when the largest computationally accessible  model space, namely 
$\mathcal{N}_{\rm max}=34\, (20)$, is employed. Not well-converged hypernuclear binding energies may impact the final conclusion about the nuclear model uncertainty in those hypernuclei. Therefore, to speed up the convergence of the NCSM calculations,  we will evolve all the employed $N\!N$, $3N$ and $Y\!N$ potentials with an SRG transformation \cite{Bogner:2006pc,Bogner:2007qb,Jurgenson:2009hq,Wirth:2019cpp,Le:2022ikc}. Like in our previous work~\cite{Le:2022ikc}, here both SRG-induced $3N$ and $Y\!N\!N$ forces are explicitly taken into account, while the SRG-induced four- and higher-body forces, whose contributions to the binding energies are expected to be small, are omitted. In most of the calculations below, we will use an SRG flow parameter of $\lambda=1.88$  fm\textsuperscript{-1} which is widely employed in both nuclear \cite{Maris:2016bia,Binder:2018pgl,LENPIC:2022cyu,Maris:2023esu} and hypernuclear calculations \cite{Wirth:2019cpp,Le:2022ikc}. At that flow parameter, a numerical uncertainty of  a  few keV can be achieved   for both  $^{\,4}_{\Lambda}\mathrm{He}$ and $^{\,5}_{\Lambda}\mathrm{He}$ for the model space $\mathcal{N}_{\rm max}=26$ and 18, respectively,  see Fig.~\ref{fig:extrapolate-E4He} (lower panels) and also Table~\ref{tab:Blambda_4_5}. It is therefore not necessary  to perform the calculations for the $A=4,\,5$  hypernuclei using  our largest computationally accessible model spaces, namely $\mathcal{N}_{\rm max} =34$ and 20, respectively.

Since we do not include any  SRG-induced interactions beyond  $3B\!F$s in the current study, it is essential  to quantify the size of 
the  possible contributions from those missing  forces to the separation energies in the $A=4,5$ hypernuclei. For that purpose, we perform  calculations for $B_{\Lambda}(^4_{\Lambda}\mathrm{He}(0^+))$ and $B_{\Lambda}(^5_{\Lambda}\mathrm{He})$ for a wide range of SRG flow parameters, namely $1.88 \leq \lambda \leq 3.0\, (4.0)$ fm\textsuperscript{-1}. The results are tabulated in Table~\ref{tab:Blambda_4_5}. Note that the separation energy $B_{\Lambda}(^4_{\Lambda}\mathrm{He}(0^+))$ at $\lambda=\infty$ (i.e. non SRG-evolved) has been computed by solving the FY equations with the bare $N\!N$, $3N$ and $Y\!N$ potentials. Overall, one observes a negligible  variation in $B_{\Lambda}(^4_{\Lambda}\mathrm{He})$ (of about $10 \pm 25 $ keV)  over the considered range of the
SRG parameter which strongly indicates that the omitted SRG-induced  forces contribute  insignificantly  to $B_{\Lambda}(^4_{\Lambda}\mathrm{He})$. In addition, there is a negligibly small difference  of  about $25 \pm 30$ keV  between the separation energies at  $\lambda=\infty$  and at a finite flow parameter ($\lambda=3.00$ fm\textsuperscript{-1}) consistent with zero within the numerical accuracy  which  again  confirms the smallness of the possible correction from  the missing induced higher-body forces to $B_{\Lambda}(^4_{\Lambda}\mathrm{He}(0^+))$. We note that  similarly small   discrepancies (about $20 \pm 20$ keV) are also observed  for the excited state separation energies $^{\,4}_{\Lambda}\mathrm{He}(1^+)$, see Table~\ref{tab:E5Helambda_NNcutoff1}. For the $^{\,5}_{\Lambda}\mathrm{He}$ 
system, we do not have the result at $\lambda=\infty$, nevertheless, with the available results,  one can still estimate a small contribution of about $100 \pm 30 $ keV from the neglected SRG-induced forces. We will see below that this inaccuracy is significantly less than the uncertainty due to the truncation of the chiral expansion.   Hence, both our numerical uncertainties and the truncated errors of the SRG evolution are sufficiently small which in turn will allow for an accurate estimate
of the theoretical uncertainties due to the underlying interactions considered in the following section.

\section{Uncertainties from the \texorpdfstring{$N\!N$ and $Y\!N$}{NN and YN} interactions} 
\label{Sec:Potentials} 

\begin{table*}[tbp]
 
\renewcommand{\arraystretch}{1.5}

\begin{center}
  \setlength{\tabcolsep}{0.3cm}
\begin{tabular}{|l|  l|    l  l|   l  l |l|}
\hline
  $\Lambda_N$  [MeV]   &   $ B_{\Lambda}(^3_{\Lambda}\mathrm{H})$ &  \multicolumn{2}{c|}{$B_{\Lambda}(^4_{\Lambda}\mathrm{He}, 0^+)$} &  \multicolumn{2}{c|}{$B_{\Lambda}(^4_{\Lambda}\mathrm{He}, 1^+)$} &  
  \multicolumn{1}{c|}{$B_{\Lambda}(^5_{\Lambda}\mathrm{He})$} \\
  & FY  & NCSM & FY & NCSM  & FY & NCSM \\
  \hline
\hline
 500\,(N$^2$LO)  &    0.118      $\pm$  0.001  &2.061  $\pm $ 0.002 & 2.06 $\pm$ 0.02 &   1.119 $\pm$  0.009  &
 1.12 $\pm$  0.02  & 3.409 $\pm$ 0.007 \\
 450\,(N$^2$LO)  &   0.125      $\pm$  0.001  &2.119  $\pm $ 0.002 & 2.13 $\pm$ 0.02   &   1.141 $\pm$  0.007  &
 1.16 $\pm$  0.02 & 3.518 $\pm$ 0.008 \\
 450\,(N$^3$LO)  &  0.122       $\pm$ 0.001 & 2.042  $\pm$  0.002 & 2.07 $\pm $0.02  &  1.07  $\pm$  0.009 &  
 1.09 $\pm$  0.02  & 3.287 $\pm$ 0.008 \\
 \hline
\hline
  400    &   0.127 $\pm$ 0.001         & 2.084  $\pm $ 0.002 & 2.12 $\pm$ 0.02  &  1.08  $\pm$   0.009 & 
 1.10 $\pm$  0.02 & 3.308 $\pm$ 0.008 \\
  450   &    0.123      $\pm$  0.001   & 2.061 $\pm 0.001$ & 2.08 $\pm$ 0.02   &  1.087  $\pm$  0.009  & 
 1.10 $\pm$  0.02 & 3.334 $\pm$ 0.008 \\
  500    &   0.118        $\pm$ 0.001  & 2.02  $\pm$ 0.001 & 2.03 $\pm$ 0.02   &  1.08   $\pm$  0.009   & 
 1.09 $\pm$  0.02 & 3.310 $\pm$ 0.008 \\
  550    &  0.113      $\pm$ 0.001  & 1.972  $\pm$  0.001 & 1.96 $\pm$ 0.02  &  1.064  $\pm$  0.009  & 
 1.07 $\pm$  0.02   & 3.245 $\pm$ 0.009 \\
\hline
\hline
\multirow{2}{*}{experiment \cite{HypernuclearDataBase} }
&  $0.164\pm 0.043$ & \multicolumn{2}{c|}{ $2.169 \pm 0.042$ \    $\quad(^4_{\Lambda}\mathrm{H})$} & \multicolumn{2}{c|}{ $1.081\pm 0.046$ \  $\quad(^4_{\Lambda}\mathrm{H})$} 
 & $3.102 \pm 0.030$  \\
 &  &  \multicolumn{2}{c|}{ $2.347 \pm 0.036$   $\quad(^4_{\Lambda}\mathrm{He})$}  &   \multicolumn{2}{c|}{ $0.942\pm 0.036$   $\quad(^4_{\Lambda}\mathrm{He})$}  &\\
\hline 
  \end{tabular}
\end{center}
\renewcommand{\arraystretch}{1.4}
\caption{\label{tab:E5Helambda_NNcutoff1} Separation energies (in MeV) for \nucl{3}{\Lambda}{H}, \nucl{4}{\Lambda}{He},  and \nucl{5}{\Lambda}{He}
 based on the
 $\mathrm{N^4LO}^+$ $N\!N$ potential with cutoffs $400-550$~MeV  
 and with inclusion of the chiral $3N$ force at N$^2$LO. For the interaction
 in the $Y\!N$ system the potential SMS $\mathrm{NLO}(550)$ is employed.
 Selected results for $\mathrm{N^2LO}$ and $\mathrm{N^3LO}$ $N\!N$ potentials 
 are included too.
}
\end{table*}

Recently performed estimates for the truncation
error of the chiral expansion for the nucleonic sector build 
primarily on approaches that do not rely on cutoff variations
\cite{Epelbaum:2014efa,Furnstahl:2015rha,Reinert:2017usi,Epelbaum:2019zqc,LENPIC:2022cyu,Maris:2023esu}. 
The cutoff dependence, or generally speaking the residual regulator dependence,
does provide a measure for the effects of high-order contributions but it is not 
a reliable tool for estimating the theoretical uncertainty due to cutoff artifacts, see also  the arguments
in Sect.~7 of Ref.~\cite{Epelbaum:2014efa}. 
Thus, in order to investigate the convergence pattern of the separation energies 
of the considered hypernuclei with increasing order, 
we have implemented the Bayesian approach of 
Refs.~\cite{Furnstahl:2015rha,Melendez:2017phj} and summarized in Appendix~1 of Ref.~\cite{Melendez:2019izc}. 
 
The calculations presented in this section utilize 
$N\!N$ potentials from LO up to N$^4$LO$^+$ and include 
the leading $3N\!F$s starting from N$^2$LO. 
However, 
they are without the leading chiral $Y\!N\!N$ interactions
and, therefore, incomplete starting from order N$^2$LO. 
We refrain from using so-called ``projected results'' (see \cite{Binder:2016bi}) which assume experimental values for certain 
binding energies arguing that these results can be fitted 
once LECs of the missing terms have been adjusted. 
We will discuss below how the missing terms might alter 
uncertainty estimates.

\subsection{Discussion of the variations} 

Let us first inspect the variation of the separation energies 
with the employed $N\!N$ potentials. 
As already mentioned in the introduction, previous bound-state calculations 
by us suggested that the $\La$ separation
energies of light hypernuclei are not very sensitive to the 
employed $N\!N$ interaction \cite{Nogga:2001ef,Haidenbauer:2019boi}.
For example, the variation of the separation energy for the 
SMS $N\!N$ potential of
Ref.~\cite{Reinert:2017usi} at order N$^4$LO$^+$ with cutoffs $\La_{\mathrm{N}}=400-550$~MeV were found to be around  
$100$~keV for $^{\,4}_{\Lambda}\mathrm{He}/^4_{\Lambda}\mathrm{H}$ \cite{Haidenbauer:2019boi}. Those for the hypertriton were in
the order of only $10$~keV. Variations of similar magnitude have been
observed in earlier calculations based 
on phenomenological interactions \cite{Nogga:2001ef}. 

 \begin{table*}[tbp]
 
 \renewcommand{\arraystretch}{1.6}
 \begin{center}
  \setlength{\tabcolsep}{0.3cm}
\begin{tabular}{|l|l|r|r|r|r|}
\hline
  & considered $N\!N$\,+\,$Y\!N$ potentials 
  & $\Delta B_{\Lambda}(^3_{\Lambda}\mathrm{H})$ 
  & $\Delta B_{\Lambda}(^4_{\Lambda}\mathrm{He},0^+)$ 
  & $\Delta B_{\Lambda}(^4_{\Lambda}\mathrm{He}, 1^+)$ 
  & $\Delta B_{\Lambda}(^5_{\Lambda}\mathrm{He})$ \\
 \hline
 \multirow{5}{*}{NCSM/FY} & N$^4$LO$^+$+N$^2$LO(550) & 3 &  43 &  44 &  45\\
  & N$^4$LO$^+$+NLO(550)     & 14 & 110 &  25 & 90 \\
   \cline{2-6}
  & N$^4$LO$^+$+LO(600)      & 25 & 194 & 223 & 970 \\
   \cline{2-6}
  & (N$^2$LO,N$^3$LO,N$^4$LO$^+$)+N$^2$LO(550) & 11 & 114 & 114 &  295 \\
 & (N$^2$LO,N$^3$LO,N$^4$LO$^+$)+NLO(550) & 14 & 147 &  88 &  273\\
 \hline
  FY \cite{Haidenbauer:2019boi}
  & N$^4$LO$^+$+NLO19(650)   & 10 & 85 &  50 &  \\
 \hline
 \hline
 Refs.~\cite{Htun:2021jnu,Gazda:2022fte} &
 variation with cutoff & 50 & 270 & 240 & 1150 \\
 Ref.~\cite{Gazda:2022fte} & standard deviation $\sigma_{\mathrm{model}}$ & 20 & 100 & 100 & 400 \\
\hline
  \end{tabular}
\end{center}
\renewcommand{\arraystretch}{1.0}
\caption{Variation ($\Delta B_{\Lambda}$) 
 of $A=3-5$ separation energies for SMS
 N$^4$LO$^+$ $N\!N$ and  N$^2$LO $3N$ potentials with cutoffs $400-550$~MeV 
 and $Y\!N$ potentials of LO, NLO, and N$^2$LO (in keV).
 (N$^2$LO,N$^3$LO, N$^4$LO$^+$) means that results for
 N$^2$LO and N$^3$LO NN potentials for selected cutoffs 
 are considered, too. }
 \label{tab:variations} 
\end{table*}

         \begin{figure*}[tbp] 
      \begin{center}
     {\includegraphics[width=0.48\textwidth,trim={0.0cm 0.00cm 0.0cm 0 cm},clip]{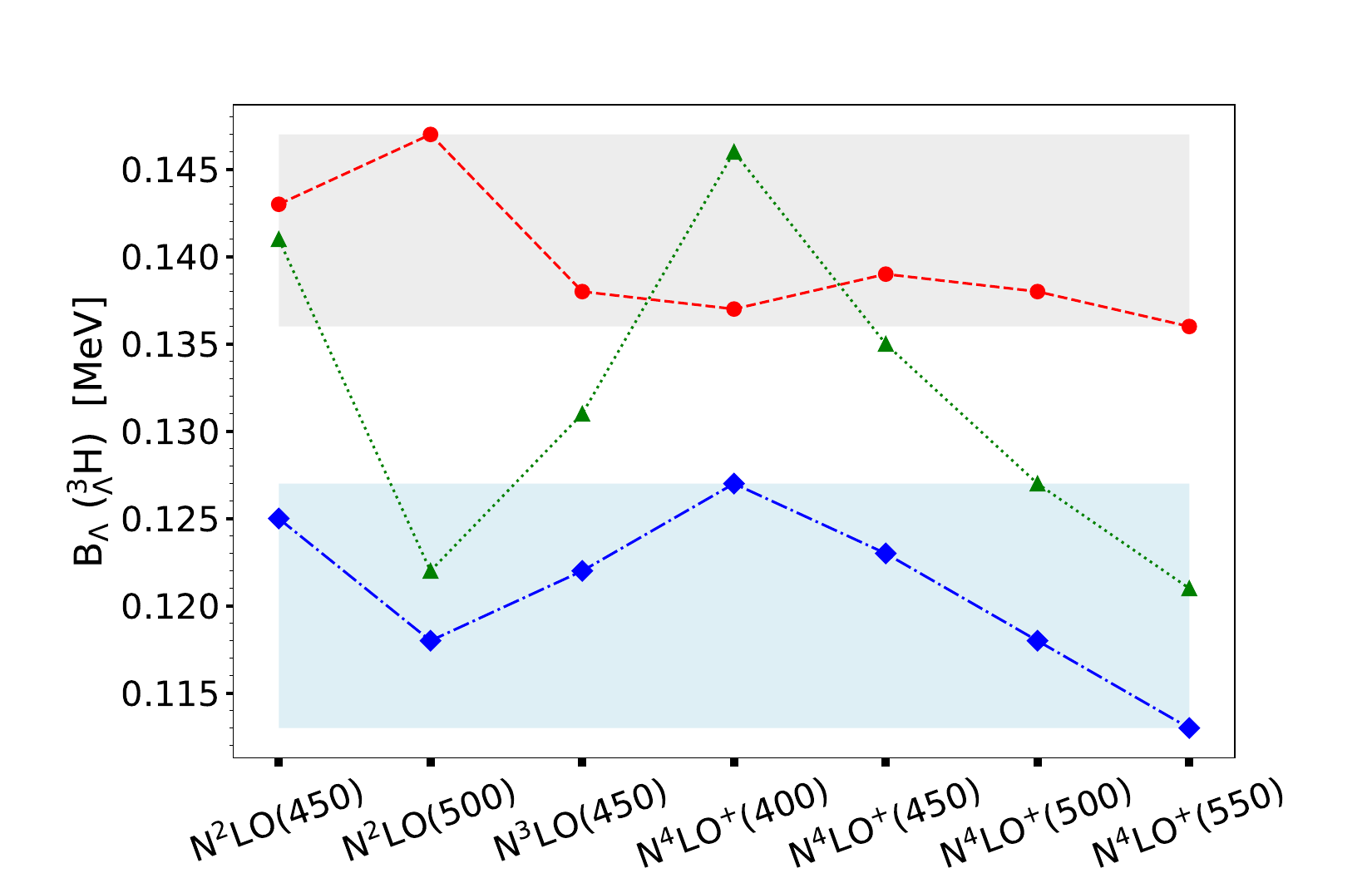}}
      {\includegraphics[width=0.48\textwidth,trim={0.0cm 0.00cm 0.0cm 0 cm},clip]{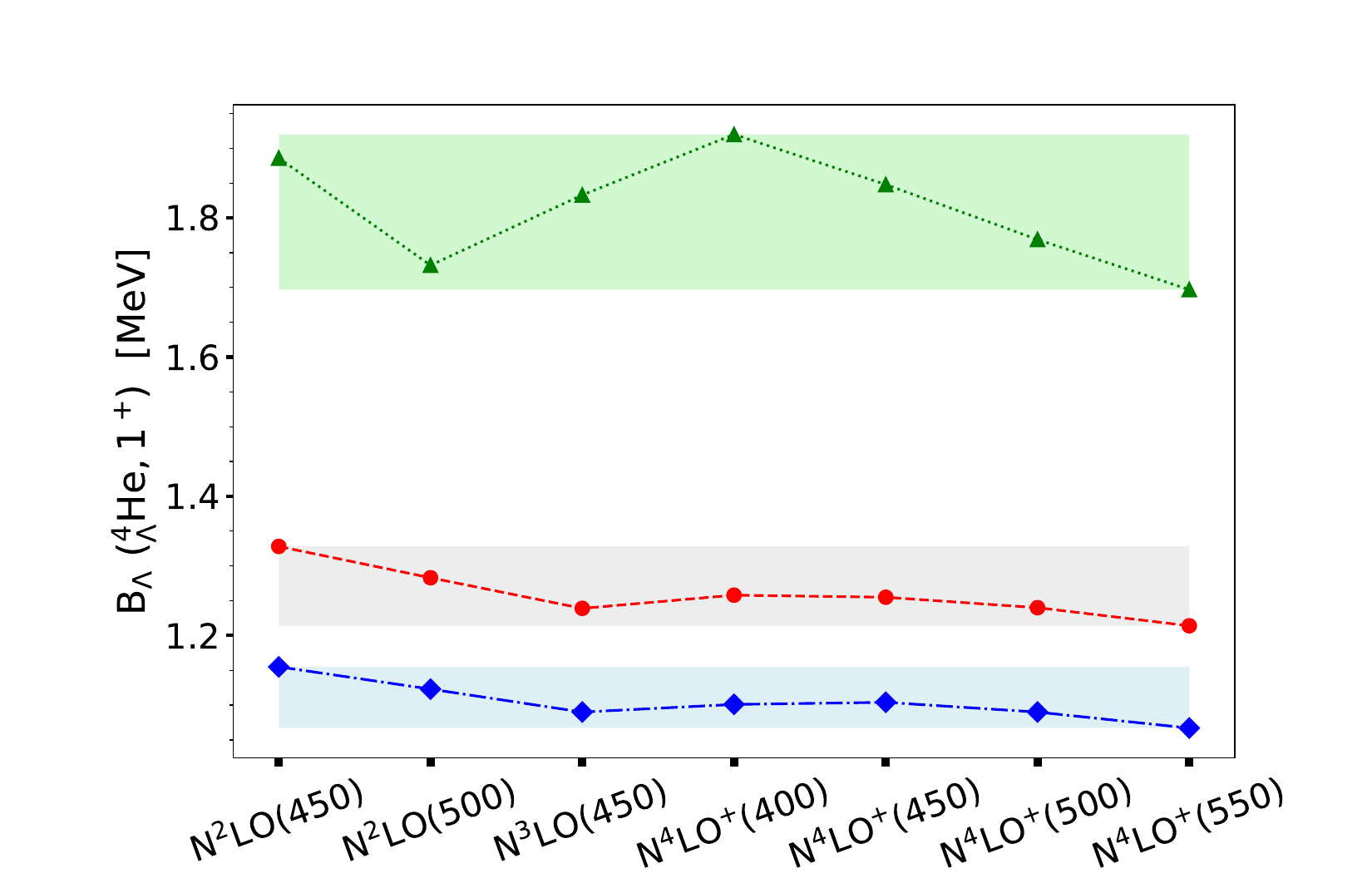}}
    {\includegraphics[width=0.48\textwidth,trim={0.0cm 0.00cm 0.0cm 0 cm},clip]{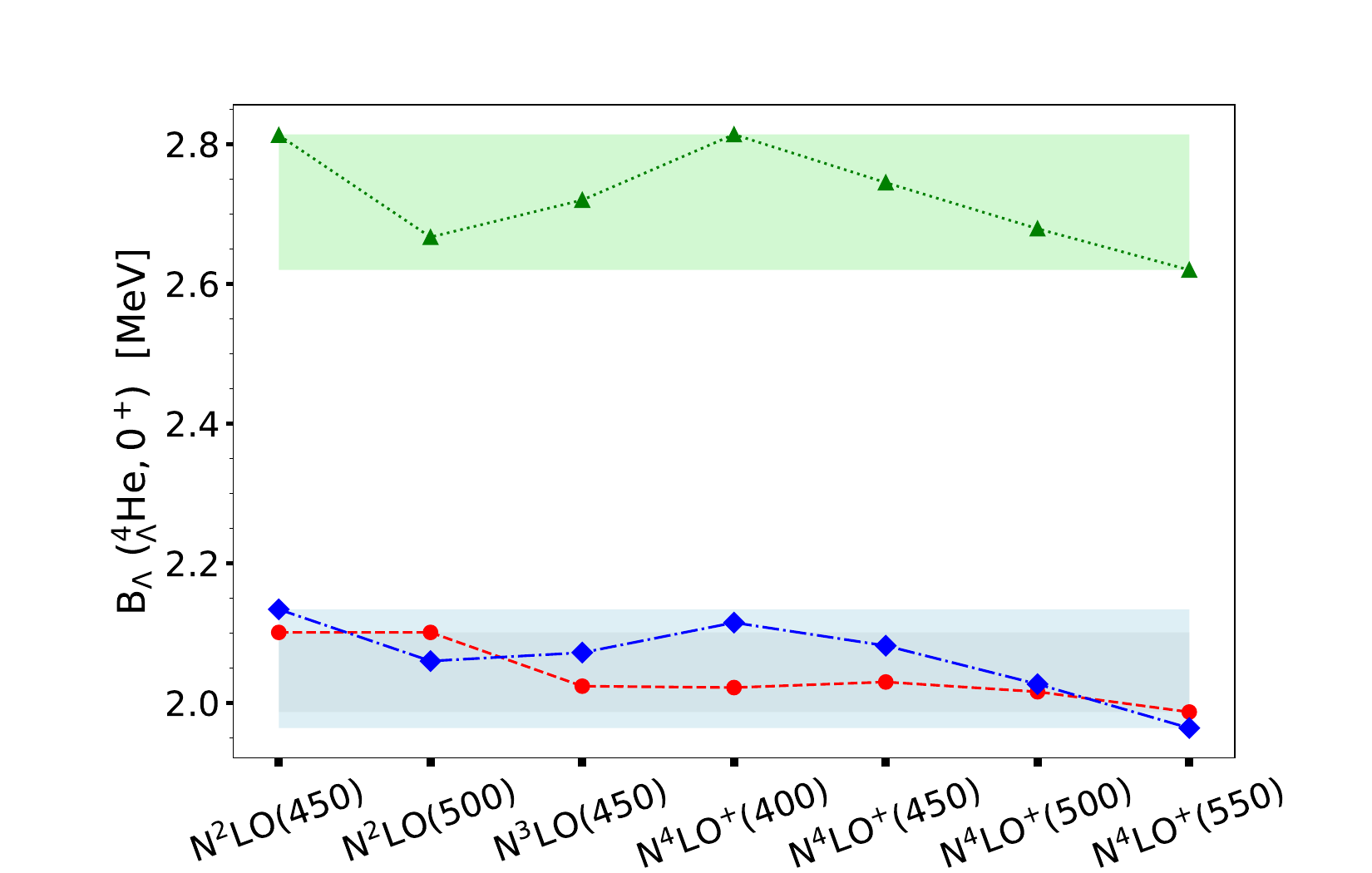}} 
     {\includegraphics[width=0.48\textwidth,trim={0.0cm 0.00cm 0.0cm 0 cm},clip]{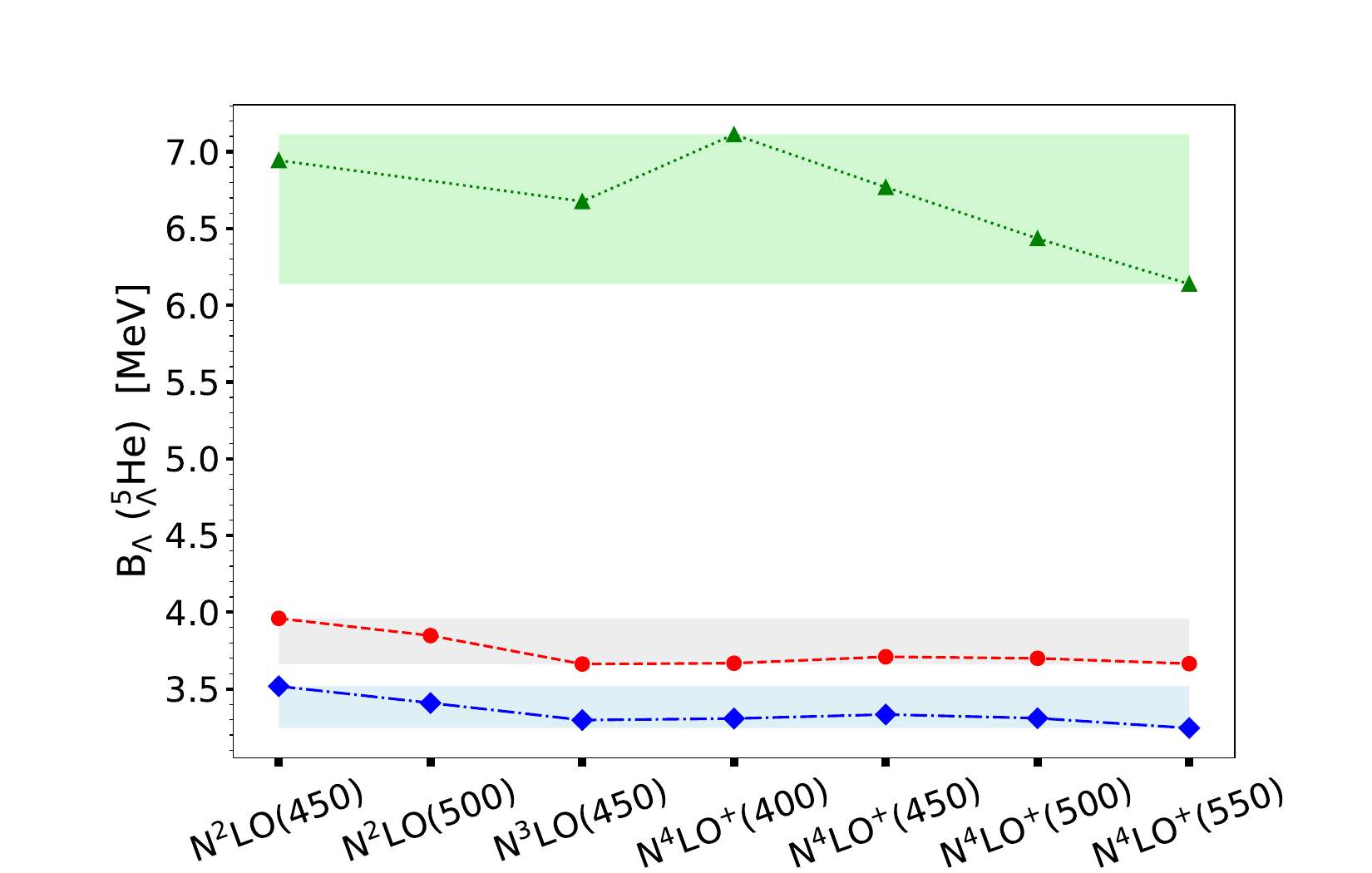}}
      \end{center}
             \caption{
             Separation energies for $^{\,3}_{\Lambda}\mathrm{H}$,
           $^{\,4}_{\Lambda}\mathrm{He}\, (0^+,1^+)$,
                 and $^{\,5}_\Lambda$He, 
                 for different combinations of $N\!N$ and $Y\!N$ interactions.
                 $Y\!N$: LO(600) (green triangles), SMS NLO(550) (blue diamonds), 
                 and SMS $\mathrm{N^2LO}$(550) (red circles). $3N\!F$s are included starting
                 from N\textsuperscript{2}LO. 
                 The employed $N\!N$ interactions are specified on the x-axis. 
                 Shaded areas show the overall variation of the separation energies for the employed  $Y\!N$ potentials at the given order.
                 The band for LO in $^3_\Lambda$H was omitted since it covers $3/4$ of the plot and obscures the size of the other bands. 
                 }
    \label{fig:nn-dep_3Hlambda}
         \end{figure*}

Separation energies for $A=3-5$ $\Lambda$ hypernuclei, 
obtained within the NCSM approach and from solving FY 
equations, are summarized in Table~\ref{tab:E5Helambda_NNcutoff1}. 
The calculations are based on the $N\!N$ and $3N$ potentials at N$^4$LO$^+$ and 
$\mathrm{N^2LO}$, respectively, with four different cutoffs. To describe 
the $Y\!N$ interaction the SMS NLO(550) potential has been employed. We 
consider also $N\!N$ potentials up to N$^2$LO and N$^3$LO with selected cutoffs   for illustration.  
As already discussed above, the small deviations between the NCSM and FY results are due to SRG induced 
four-baryon interactions that are omitted in the calculations. 
It sticks out that the numerical uncertainty 
of the FY results is similar or even larger than the deviation to the NCSM results. Therefore, below 
we will estimate our uncertainties based on the NCSM results if available. A
 graphical representation of the separation energies is provided in
Fig.~\ref{fig:nn-dep_3Hlambda}. Here, in addition,  results for the 
SMS $\mathrm{N^2LO}(550)$ $Y\!N$ potential as well as  some results for the LO(600) $Y\!N$  potential are shown.

One can see from Table~\ref{tab:E5Helambda_NNcutoff1} that the overall variation 
of the $^{\,5}_{\Lambda}\mathrm{He}$ separation energy is indeed very small.
Furthermore, in general, the variations are smaller than the difference to the
experimental value. The latter difference will eventually be accounted for via 
inclusion of $Y\!N\!N$ $3B\!F$s and/or with improved $Y\!N$ interactions.
The situation for the $^{\,4}_{\Lambda}\mathrm{He}$ separation energies is
similar. Also here for the $0^+$ as well as for the $1^+$ state, the variations
due to the employed $N\!N$ potential are small and specifically smaller than
the difference to the empirical separation energies. Note that, since we do not include charge symmetry breaking potentials in the current study, our results for $B_{\Lambda}(^4_{\Lambda}\mathrm{He})$ should be compared to the experimental values for both $^{\,4}_{\Lambda}\mathrm{He}$ and  $^{\,4}_{\Lambda}\mathrm{H}$ hypernuclei.

A detailed overview of the variation of the separation energies
for the considered hypernuclei 
due to different chiral orders and different cutoffs $\Lambda_{N}$ of the underlying $N\!N$ potentials is provided in Table~\ref{tab:variations},
see also Fig.~\ref{fig:nn-dep_3Hlambda}. For the set of N$^4$LO$^+$ $N\!N$
potentials combined with the NLO or N$^2$LO $Y\!N$ 
interactions the variations of $B_{\Lambda}(^5_{\Lambda}\mathrm{He})$
are $45-90$~keV. 
They increase to $295$~keV when $N\!N$ potentials of lower order are
considered additionally. With regard to $B_{\Lambda}$ of the $^{\,4}_{\Lambda}\mathrm{He}$ $1^+$ state, the variation for the N$^4$LO 
set is only of the order of $25-44$~keV and increases to $114$~keV 
when N$^2$LO/N$^3$LO $N\!N$ interactions are taken into account. 
Concerning the $0^+$ state, the variation for the N$^4$LO$^+$ set is 
$43-110$~keV. It becomes slightly larger but remains 
of similar magnitude by considering lower order $N\!N$ interactions.
Clearly, using the most sophisticated (and most accurate) $N\!N$ interactions significantly reduces the sensitivity of $B_{\Lambda}$ to the employed  $N\!N$ potentials. In passing, let us also mention that, as shown  in Ref.~\cite{LENPIC:2022cyu}, by including the higher-order corrections to the $N\!N$ potentials
up through fifth order, the systematic overbinding observed in nuclei with $A > 10$  reported in their earlier study in Ref.~\cite{Maris:2020qne} when only $N\!N$ and $3N$ interactions at $\mathrm{N^2LO}$ were employed, is practically resolved. 

In order to compare the variations found by us with the ones reported by
Gazda et al., we simply digitized their results from Figs.~3 and 7 of Refs.~\cite{Htun:2021jnu,Gazda:2022fte}, respectively. The 
corresponding values are also listed in Table~\ref{tab:variations}. 
Actually, those authors considered variations due to the cutoff as well as
variations due to the fitting region. The values we provide in the table 
are an average over those for different fitting regions. 
Obviously, the variations observed in that study are 
about 3 times larger for $^{\,4}_{\Lambda}\mathrm{He}\, (0^+)$, and 
practically a factor 10 larger for $^{\,4}_{\Lambda}\mathrm{He}\, (1^+)$ 
and $^{\,5}_{\Lambda}\mathrm{He}$ than the variations we find for the 
NLO and N$^2$LO $Y\!N$ potentials in combination with N$^4$LO$^+$ 
$N\!N$ potentials. 
The variation for $B_{\Lambda}(^3_{\Lambda}\mathrm{H})$
is likewise a factor 3 larger than ours. 
However, when we use a LO $Y\!N$ potential (see the corresponding line 
in Table~\ref{tab:variations} and Fig.~\ref{fig:nn-dep_3Hlambda}) the 
variations become comparably large as those reported 
in~\cite{Htun:2021jnu,Gazda:2022fte}. 
%

We identified three possible sources for the differences in the 
variations. First, 
we expect a sensitivity to the actual size of the separation energies. 
In general, the LO $Y\!N$ interactions overbind the considered 
hypernuclei substantially (see Fig.~\ref{fig:nn-dep_3Hlambda} and also 
Tables I, II in \cite{Gazda:2022fte}), and naturally, a significantly 
larger variation is expected .
For $^{\,3}_\Lambda \mathrm{H}$, the value of $B_{\Lambda}$
is fixed by construction, for all considered $Y\!N$ potentials. 
Since here we observe an increased dependence on the $N\!N$ interaction too, when a LO $Y\!N$ potential is used, we believe that the lack of 
short-range repulsion in the LO $Y\!N$ potentials 
is also a potential source for the difference 
to the results with NLO and N$^2$LO. 
This deficiency of the LO interactions can lead 
to an increased 
sensitivity to the details of the short-range part of the $N\!N$ interactions. 
Third, there is presumably an effect from the employed regularization scheme. 
The N$^2$LO$_{sim}$ potentials employed by Gazda et al. build on a non-local 
regulator for all components of the interaction. The SMS $N\!N$ potentials 
by Reinert et al. are based on a novel regularization scheme where a local regulator 
is applied to the pion-exchange contributions and only the contact terms, 
being non-local by themselves, are regularized with a non-local function.
As discussed thoroughly in \cite{Epelbaum:2014efa,Reinert:2017usi}, 
a local regulator for pion-exchange contributions leads to a reduction of 
the distortion in the long-range part of the interaction and, thereby,
facilitates a more rapid convergence already at low chiral orders. 
This affects predominantly $P$- and higher partial waves. 
In this context note that the optimal cutoff range
is shifted from $450-600$~MeV (Carlsson et al.
\cite{Carlsson:2015vda}) to $400-550$~MeV
(Reinert et al \cite{Reinert:2017usi}).

Finally, and for clarification, we want to emphasize that the ``model 
uncertainties'' quoted in the abstract and in the summary of 
Ref.~\cite{Gazda:2022fte}, 
have been deduced from the variance $\sigma^2$(NNLO$_{\rm sim}$) as
specified in Eq.~(33) of that paper. As for reference those uncertainties
are listed in the last line of Table~\ref{tab:variations}. It is 
important to stress that those values, which are noticeably smaller than 
the variations, cannot and should not be compared with our results. 

\subsection{Estimate for the truncation error}
\label{Sec:Potentials2}

Given that we have results for different orders of the chiral $N\!N$ and $Y\!N$ 
interactions at our disposal, we are now able to perform a more complete 
analysis of the uncertainties due to truncation in the chiral expansion. 
To this aim, we follow the Bayesian 
approach of \cite{Melendez:2017phj} and Ref.~\cite{Melendez:2019izc}, cf.
the appendix on the pointwise model. Assume that the observable $X$ (here the separation energies) also follow the power counting of the potential. If 
the chiral expansion is truncated at order $K$, the observable $X_{K}$ and the corresponding truncation error $\delta X_{K}$ can be expressed as
\begin{equation}
    X_{K}^{} = X_{\rm ref}^{} \left( \sum_{k=0}^{K} c_k  Q^k \right); \,\,    \delta X_{K} = X_{\rm ref} \left( \sum_{k=K+1}^{\infty} c_k  Q^k \right) .
\end{equation}
where $Q = M_{\pi}^{\rm eff} / \Lambda_b  $ is the chiral EFT expansion parameter  
\cite{Epelbaum:2019wvf},  $X_{\rm ref}$ is a dimensionful quantity that sets the overall scale and $c_{k}$ are the dimensionless expansion coefficients.  The expansion parameter is given by an effective pion mass $M_{\pi}^{\rm eff} $ and the breakdown scale $\Lambda_b $. 
The expansion coefficients $c_{k}\, (k =0,2, \cdots K)$
are obtained from the separation energies computed at two consecutive orders, \\
$c_{k+1} = (B_{\Lambda}^{(k+1)} - B_{\Lambda}^{(k)})/ ( Q^{k+1} X_{\rm ref})$. 
In order to obtain the posterior probability distribution for the truncation error $\delta X_{K}$ based on our knowledge of the coefficients  $c_{k}\, (k =0,2, \cdots K)$, we further assume that all the expansion coefficients are independently and identically distributed (iid). 
The priors  follow the ``pointwise'' distribution given in Eq.~(A2) in the appendix of
Ref.~\cite{Melendez:2019izc}, namely a normal distribution with 
variance $\overline{c}^2$
\begin{equation}
    c_{k} | \overline{c}^2   \stackrel{\rm iid}{\sim} \mathcal{N}(0,
     \overline{c}^2)
\end{equation}
where the distribution of $\overline{c}^2$ follows an inverse 
$\chi^2$ distribution 
\begin{equation}
 \overline{c}^2 \sim \chi^{-2} (\nu_0, \tau^2_{0})
\end{equation}
depending on the two hyperparameters $\nu_0$ and $\tau^2_0$. 
The analytical expression for the posterior distribution of the truncation error $\delta X_{K}$ is also given in
Eq.~(A12) of the same appendix. 

Clearly,  the truncation error will be contingent on the expansion
parameter $Q$ as well as on  our choices for the parameters $\nu_0$ and $\tau^2_0$. We  have compared results for 
non-informative priors (not preferring any maximal value of $c_k$, i.e. using the parameter 
$\nu_0=0$) and more informative priors for $c_k$ with $\nu_0>0$. It turns out that the estimated truncation errors  were not very sensitive  
to this parameter, therefore, we finally chose $\nu_0=1.5$ which was also used in \cite{LENPIC:2022cyu} for 
studying the convergence of calculations of ordinary nuclei. We also followed \cite{LENPIC:2022cyu}  and selected 
$\tau_{0}^2=2.25$.   

\begin{figure*}[tbp]
    \centering
    \includegraphics[width=0.4\textwidth]{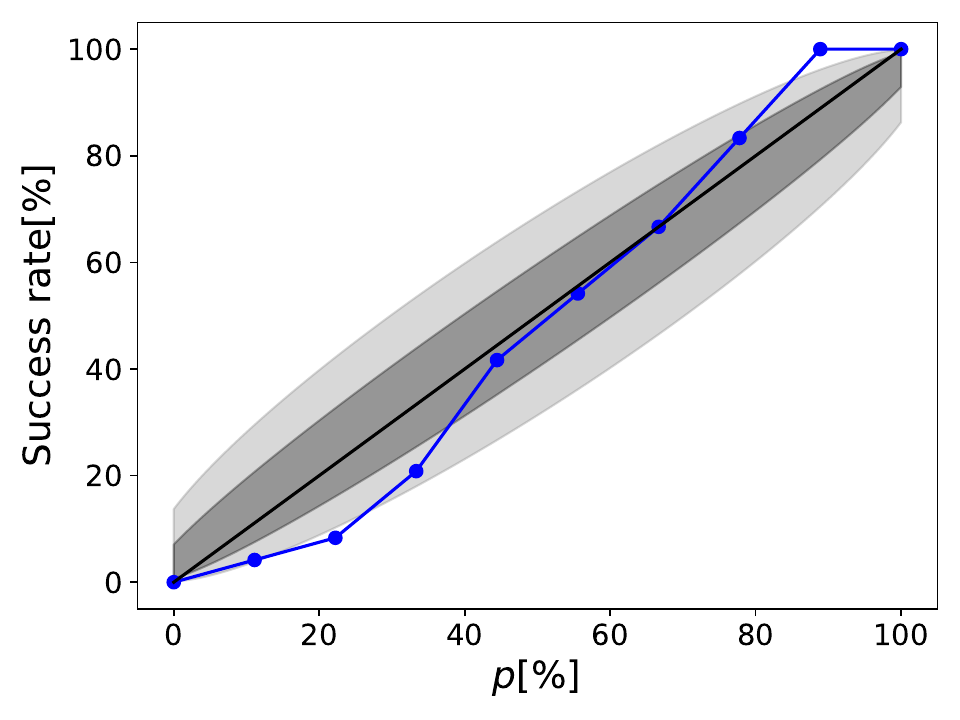}
    \includegraphics[width=0.4\textwidth]{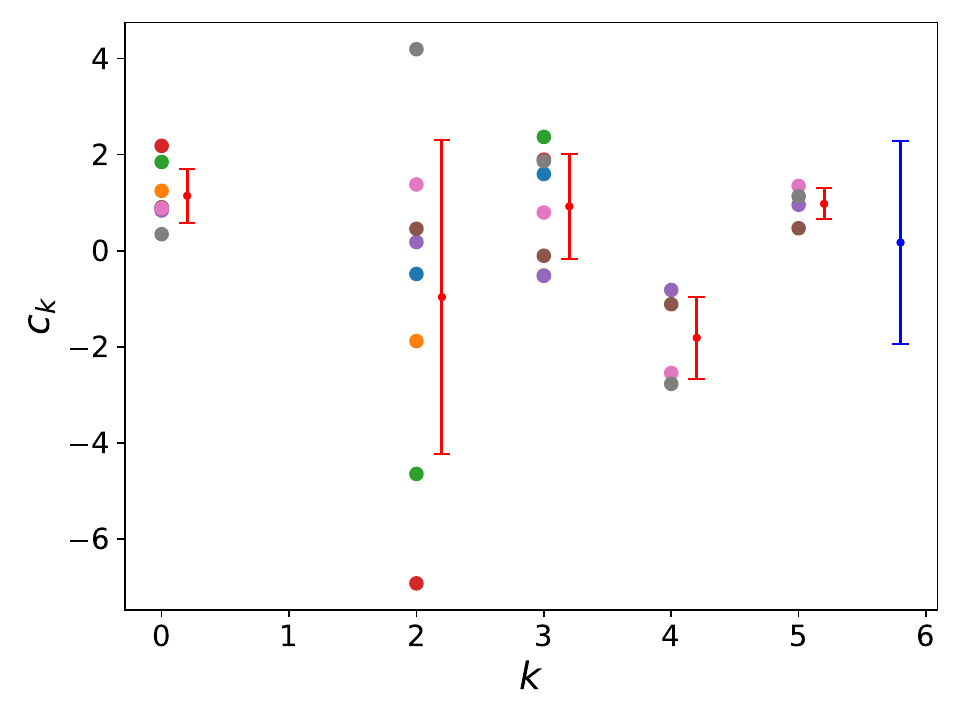}
    \caption{\label{fig:consistencnnyn}(Left) Consistency plots for comparing the actual 
    changes in higher orders to the expected values for  $p[\%]$ DoB intervals.
    (Right) Values of the $c_k$ coefficients extracted using the corresponding experimental separation energies as the reference value. Also shown are the average and standard variation order by order and in total.
    }
\end{figure*}

To learn the expansion parameter $Q$, we assume a normal distribution for the prior  of $Q$   and use the expression in Eq.~(A19) of Ref.~\cite{Melendez:2019izc} to compute the posterior distribution of $Q$. 
In order to  increase the statistics, when learning the expansion parameter $Q$ we employ the combined data that contains all the separation energies for $A=3-5$ hypernuclei computed at different orders of $N\!N$ and of  $Y\!N$ interactions. We  obtained the value of  $Q=0.4$, which 
is slightly larger than the one used for  ordinary nuclei \cite{LENPIC:2022cyu}. This could probably be related to the 
small number of our data that are available for determining the distributions. 
We also test the validity of our choice for $Q$ by
generating consistency plots as proposed in Ref.~\cite{Melendez:2017phj}
that show the comparison between the obtained rates of the overlap of higher-order calculations with lower-order degree of believe (DoB) intervals and the expected values, see Fig.~\ref{fig:consistencnnyn}.
Clearly, with the chosen value of $Q =0.4$, our uncertainty estimates are statistically consistent with the observed changes due 
to higher-order contributions. Finally, let us remark that we also extracted the $Q$ parameter using two other sets of data, referring to as the $N\!N$  and $Y \!N$ convergence studies. In the former case, the data are composed 
of  separation energies at various $N\!N$ orders ($k=0,2,3,4,5$) but with a fixed $Y\!N$ order. In the latter case, we used the data set that consists 
of the energies 
at a fixed $N \!N $ order but for different $Y \!N$ orders ($k=0,2,3$).  The obtained $Q$ values in the two cases are in general consistent with each other and with the result obtained when the combined data is used.   
The most relevant 
difference between the $N\!N$ and $Y\!N$ convergence studies  was probably the fact that the  expansion parameter $Q$ tends to be smaller for 
$N \!N$   than for $Y \!N$. The deviation was, however, small enough so 
that a combined-data analysis is preferred because of its higher statistics. Therefore, we will use the expansion parameter  $Q=0.4$ for estimating the truncation errors.

The obtained distribution of $c_k$ coefficients is also interesting. Their dependence on the 
order $k$ of the expansion is shown on the right panel  of Fig.~\ref{fig:consistencnnyn}
together with the average values per order and the complete average with standard deviation. 
For their extraction, we chose reference values close to the corresponding experimental separation energies
 in order to be independent of the LO result. The latter  might be altered
by choosing a quite small singlet scattering length in order to match the \nucl{3}{\Lambda}{H}
separation energy \cite{Polinder:2006eq}. 
In addition, because of this choice for $X_{\rm ref}$, we are able to use all coefficients  for 
determining the posteriors.  We stress that  the final truncation errors  are  independent of the  reference value.
Interestingly, the NLO coefficients have a tendency to be larger than all the other ones.  This tendency is also observed for the expansion  coefficients obtained for light nuclei \cite{LENPIC:2022cyu}.  Overall, 
all the expansion coefficients are  however of natural size and, therefore, the value of   $Q=0.4$ for the expansion scale  seems to be consistently chosen. 
For this extraction, it has been assumed that the differences of the higher-order contributions are of the order naively expected. We have also attempted to 
analyse the results assuming that the expected corrections at $\mathrm{N^2LO}$ and at
higher orders are of the order $k=3$ because the chiral $Y\!N\!N$ forces contributing at this order is missing.  In this case, the higher-order $N\!N$ expansion coefficients become unnaturally small. This in turn 
supports our assumption that these differences are indeed of the expected order. 
Note that this assumption is however not true for the regulator dependence which will ultimately 
be counterbalanced by a $Y\!N\!N$ $3B\!F$ once it has been taken into account. The
cutoff dependence is therefore a $Q^3$ effect for NLO and all higher orders.

\begin{figure*}[tbp]
    \centering
    \includegraphics[width=0.8\textwidth]{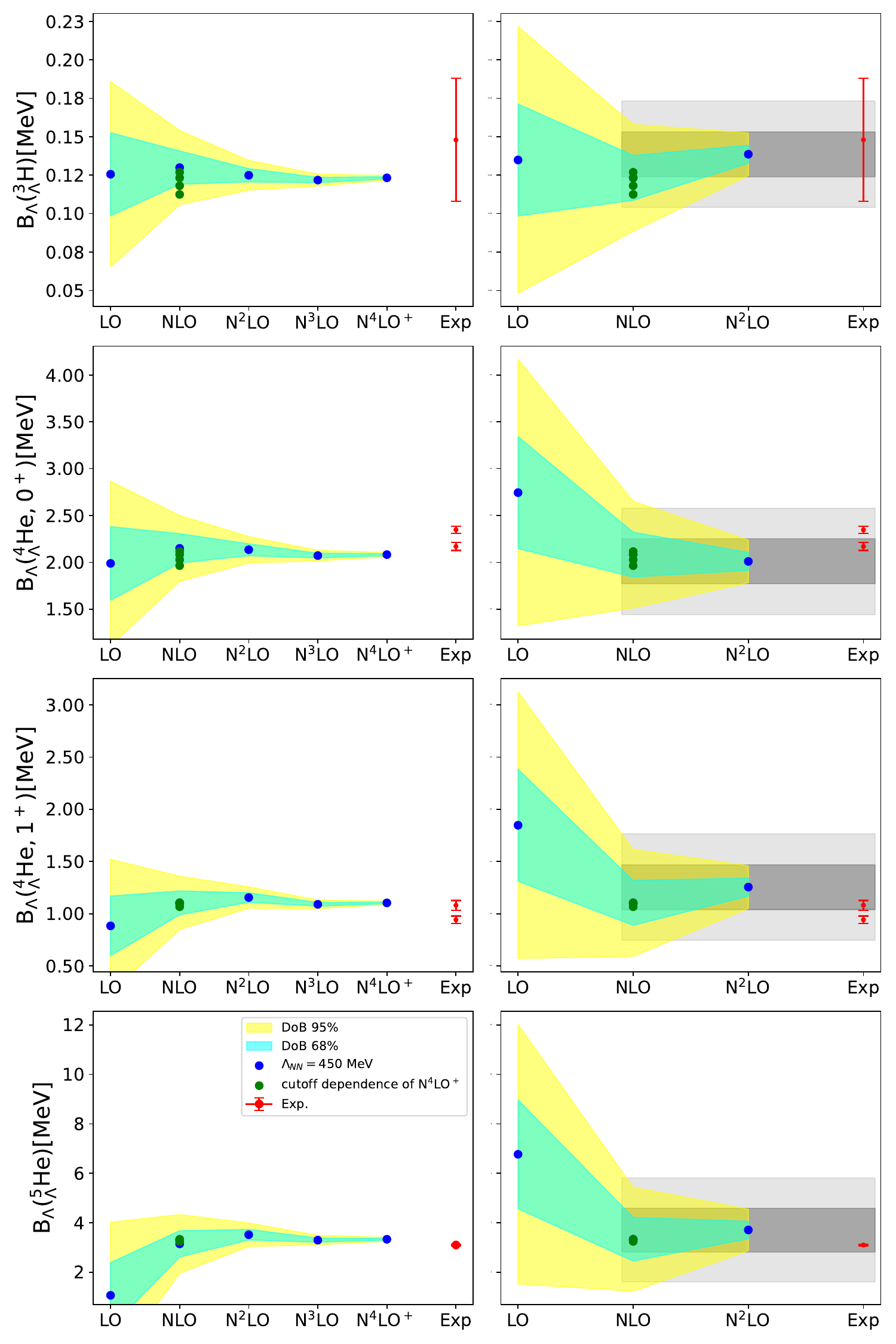}
        
    \caption{Comparison of the convergence with respect to the chiral order of 
    the employed $N\!N$ (left) and  $Y\!N$ (right) potentials  for  
     \nucl{3}{\Lambda}{H}, \nucl{4}{\Lambda}{He($0^+$)}, \nucl{4}{\Lambda}{He($1^+$)}and \nucl{5}{\Lambda}{He} (from top to bottom). \label{fig:compnnyn}}
    
\end{figure*}

Having the hyperparameters and the expansion scale fixed, we are now at the position to analyze the convergence pattern of the separation energies with respect to chiral order. As already 
documented in the previous section, the separation energies 
are much less sensitive to the $N\!N$ than to the $Y\!N$ interaction. This is also manifest 
in a different size of $c_k$ coefficients for the $N\!N$ convergence and the $Y\!N$ convergence.
We therefore analyze the convergence and extract the  truncation uncertainties  for $N\!N$ and $Y\!N$ separately. The convergence of the separation energies for $A=3-5$ hypernuclei with respect to $N \!N$ and $Y \!N$ orders are shown in the left and right panels in  Fig.~\ref{fig:compnnyn}, respectively. The bands show the expected truncation errors at each orders. Clearly, the large expansion parameter 
leads only to a slow decrease of this uncertainty at higher orders. 
It clearly sticks out that  
the variation due to the $N\!N$ interaction is much smaller than the one due to the 
$Y\!N$ interaction.  In order to compare the $N\!N$ cutoff variation 
with the relevant uncertainty estimate, we include also results for different 
$N \!N$ cutoffs,  see green points. Although these calculations were performed 
at order N$^4$LO$^+$, we show them in the figure at NLO since the cutoff 
variation will be ultimately mostly observed by the only N$^2$LO contribution 
that we are not taking into account, namely the leading $Y\!N\!N$ $3B\!F$. 
As can be seen, the $N\!N$ cutoff variation is consistent with  but in most of the case smaller than the 68\% DoB interval.   This is consistent with our observation in the previous section and with  the general expectation that 
the  cutoff variation as well as  the 
dependence on the chiral order of the $N \!N$ interaction is of less relevance when predicting  $\Lambda$ separation energies.

The most dominant  uncertainty is due to the truncation of the chiral 
expansion of the $Y\!N$ interaction, as can be clearly seen in the right panel in Fig.~\ref{fig:compnnyn}. Here the grey 
bands indicate the uncertainty at NLO attached to the result at order N$^2$LO. 
This is the relevant quantity for the comparison to the experimental separation energies shown in red symbols
since all calculations do not include the leading chiral $Y\!N\!N$ $3B\!F$. 
Note that both experimental separation energies of 
\nucl{4}{\Lambda}{H}  and \nucl{4}{\Lambda}{He} are included in the figure because
our calculations have been performed with isospin conserving 
interactions that cannot properly predict the charge symmetry 
breaking differences of the separation energies of these mirror 
hypernuclei. It can be seen that all experimental energies 
are within the 68\% DoB intervals. The NLO uncertainties are substantial and 
significantly larger than the experimental uncertainties for $A=4$ and $5$. Only 
for \nucl{3}{\Lambda}{H}, the experimental and theoretical uncertainty 
are comparable, justifying our choice to constrain the strength of 
the $Y\!N$ interaction in the $^1S_0$ partial wave 
by the \nucl{3}{\Lambda}{H} separation energy \cite{Haidenbauer:2013oca,Haidenbauer:2023qhf}. 

\begin{table}[tbp]
 \renewcommand{\arraystretch}{1.6}
    \centering
      \setlength{\tabcolsep}{0.4cm}
    \begin{tabular}{|l|c c|}
    \hline
     nucleus & $\Delta_{68}(N\!N)$ & $\Delta_{68}(Y\!N)$ \\
     \hline 
       \nucl{3}{\Lambda}{H}           &  0.01 & 0.02 \\
       \nucl{4}{\Lambda}{He\,$(0^+)$}   &  0.16 & 0.24 \\
       \nucl{4}{\Lambda}{He\,$(1^+)$}   &  0.11 & 0.21 \\
       \nucl{5}{\Lambda}{He}          &  0.53 & 0.88 \\
      \hline 
    \end{tabular}
    \caption{Half the size of the 68\% DoB intervals for the $\Lambda$ separation energy 
           at NLO based on the convergence with respect to the 
           $Y\!N$ and $N\!N$ interactions (in MeV).}
    \label{tab:dobynnn}
     \renewcommand{\arraystretch}{1.0}
\end{table}

In order to obtain an estimate of the size of the missing $Y\!N\!N$ force contributions,
 we have summarized half the size of the NLO 68\% DoB 
interval in Table~\ref{tab:dobynnn} for both, the $N\!N$ and the $Y\!N$ convergence. 
The dependence on the $N\!N$ interaction is generally a factor of two smaller than the one 
on the $Y\!N$ interaction. It is however larger than the one anticipated from older 
calculations comparing results for different phenomenological $N\!N$ interactions 
\cite{Nogga:2001ef}. Incidentally, the values are roughly in line with 
the ``model uncertainties'' from Ref.~\cite{Gazda:2022fte} that were 
based on averaging of the interactions dependence. As discussed in the 
previous subsection, the true dependence on the $N\!N$ interaction 
is actually larger, c.f. Table~\ref{tab:variations}.

 The relevant quantity for assessing the size of the $Y\!N\!N$ $3B\!F$
is the NLO 68\% DoB for $Y\!N$ since this quantity is larger. The $3B\!F$ contribution for the hypertriton is estimated to be roughly 15 keV. It is compatible with the result of 
a first explicit (though incomplete) evaluation of $3B\!F$s for $^{\,3}_\La\mathrm{H}$ 
by Kamada et al.~\cite{Kamada:2023txx}, which suggests a contribution of around 20 keV. In that work
only the contribution due to $2\pi$-exchange has been taken into account. We consider the nice agreement as a confirmation for the procedure we follow. In any case, it is important to note that 
the $3B\!F$ effect on
the hypertriton separation energy is found/ estimated to be smaller than the experimental uncertainty.

For $A=4$, the $Y\!N\!N$ $3B\!F$ can be expected to contribute in 
the order of $200$~keV. Also this estimate is in line 
with previous results. In Ref.~\cite{Haidenbauer:2019boi}, we observed that the NLO13 and NLO19 $Y\!N$ potentials 
exhibit a regulator dependence of up to $210$~keV and
variations of the separation energies of up
to $320$~keV due to dispersive effects associated with the 
$\Lambda N$-$\Sigma N$ coupling   
which both can be taken as estimate for $Y\!N\!N$ $3B\!F$
contributions. 
The estimate here, based on the convergence pattern of the chiral 
expansion, is of similar size. 
For \nucl{5}{\Lambda}{He}, the comparison of NLO19 and NLO13 can 
again provide hints to the size of $3B\!F$ effects. We found in 
Ref.~\cite{Le:2022ikc} that 
the result for NLO13 and NLO19 differs by $1.1$~MeV which gives a lower bound 
of possible $Y\!N\!N$-force contributions. Therefore, also the estimate in Table~\ref{tab:dobynnn}
of $900$~keV appears to be reasonable.  

Additionally, we employed the approach proposed by Epelbaum, Krebs and
Mei{\ss}ner (EKM) \cite{Epelbaum:2014efa} for estimating the uncertainty as outlined in the appendix. This estimated error  depends strongly on the expansion parameter chosen. It turns out that for standard values of $Q=0.31$, the estimates are well in line with the Bayesian results. For $Q=0.4$, the EKM estimates are somewhat larger but  still of similar order as the statistically motivated ones. 

It is also interesting to look at the prospective N$^2$LO uncertainties once the 
leading $Y\!N\!N$ interactions are included. In our analysis, we find 6, 100 and 350~keV 
for the A=3, 4 and 5 hypernuclei, respectively. These estimates are however strongly dependent 
on the expansion parameter $Q$. For example, for $Q=0.3$ as 
in \cite{LENPIC:2022cyu}, 
we find N$^2$LO uncertainties of 3, 50 and 200~keV. 

\section{Summary} 
\label{Sec:Summary} 
In this work, we have investigated various aspects relevant
for the theoretical uncertainties of calculations of
separation energies of $\Lambda$ hypernuclei with $A \le 5$.
These light hypernuclei have attracted some attention recently because 
their properties are mostly determined by the $S$-wave $Y\!N$ interactions 
which are reasonably well constrained by the available $Y\!N$ data and the 
hypertriton separation energy. 
To a great extent the effort for providing a quantitative assessment of 
the uncertainties of our few-body calculations was  
motivated by the study of Gazda et al. \cite{Gazda:2022fte} which suggested 
that even the employed $N\!N$ ($3N$) interactions might have an significant impact 
on the uncertainty of the predicted hyperon separation energies. 

In the present work, we considered two possible sour\-ces for uncertainties. 
First, there is the 
numerical uncertainty which, in our case, is caused by discretization and/or 
truncation of the model space in the no-core shell model calculation,
and also due to neglected contributions of 
SRG-induced four- and more-baryon interactions. 
By comparing two extrapolation methods and benchmarking to results
from FY calculations, we found that 
the numerical uncertainties are well under control and are 
actually irrelevant in comparison to other effects. 
The other source of uncertainties considered are
differences in the employed $N\!N$ (plus $3N$) 
and $Y\!N$ potentials. 
Our results for the hyperon separation energies do show some 
dependence on the underlying $N\!N$ interaction. 
However, compared to Gazda et al.~\cite{Gazda:2022fte}, the variations
are considerably smaller. 
A detailed analysis of our calculations suggests that the 
significant reduction is very likely due to the use of higher order 
$Y\!N$ interactions and of higher order $N\!N$ interactions. 
In fact, the effects due to truncating the chiral order of the $Y\!N$ 
interaction are the larger and most relevant ones 
and have been quantified in this work for the first time. 
It should be said that the way how regularization is implemented 
(all non-local or semi-local) could play a role, too, though on a 
less significant level. 

Altogether, it is reassuring to 
observe that our NLO and (incomplete) N$^2$LO results agree with the 
experimental separation energies within the estimated NLO 
truncation error. 
They show that it is now of high 
importance to also include the missing chiral $Y\!N\!N$ 
three-body force that starts contributing 
at order N$^2$LO. Work in this direction is in progress. 
The present calculation indicates that their contribution is needed
and can lead to a consistent and accurate description of all 
$s$-shell hypernuclei. 

Independently, it is important to get more experimental input to
facilitate a better determination of the $Y\!N$ interaction. 
Indeed, in the future more extensive data on the $\La p$ system, 
i.e. angular distributions and possibly polarizations,
should become available thanks to the J-PARC E86 experiment
\cite{Miwa:2022coz}.
A major advantage of the 3N system is that there the underlying
$N\!N$ interaction can be examined also via $Nd$ scattering 
and/or break-up observables. 
Some of the observables accessible in this way are known to be not 
very sensitive to the $3N\!F$  and, thus, provide an excellent direct 
and reliable testing
ground for the properties of the $N\!N$ potentials.  
Unfortunately, so far, for $\Lambda d$ scattering, we have neither
data nor calculations based on modern $Y\!N$ interactions. 
However, there are plans for measurements of $\Lambda d$ scattering at 
JLab \cite{Tumeo:2021} and experimental studies of the $\Lambda d$ 
correlation function \cite{Haidenbauer:2020uew} are under way at CERN 
by the ALICE Collaboration \cite{ALICE:2020fuk}. Finally, also a $\La nn$ 
resonance \cite{HallA:2022qqj} would provide an important additional
constraint, though its existence is still under debate.

\begin{acknowledgements}
This project is part of the ERC Advanced Grant ``EXOTIC'' supported the European Research Council (ERC) under the European Union’s Horizon 2020 research and innovation programme (grant agreement No. 101018170). This work is further supported in part by the Deutsche Forschungsgemeinschaft (DFG, German Research Foundation) and the NSFC through the funds provided to the Sino-German Collaborative Research Center TRR110 ``Symmetries and the Emergence of Structure in QCD'' (DFG Project ID 196253076 - TRR 110, NSFC Grant No. 12070131001), the Volkswagen Stiftung (Grant No. 93562)
and by the MKW~NRW under the funding code NW21-024-A.
The work of UGM was supported in part by The Chinese Academy
of Sciences (CAS) President's International Fellowship Initiative (PIFI)
(grant no.~2018DM0034). We also acknowledge support of the THEIA net-working activity 
of the Strong 2020 Project. The numerical calculations were performed on JURECA
of the J\"ulich Supercomputing Centre, J\"ulich, Germany.
\end{acknowledgements}

\appendix 
\section{Uncertainty estimate following EKM}

For estimating the truncation error of the chiral expansion
we also applied the EKM approach \cite{Epelbaum:2014efa}.
The concrete expression used to calculate an uncertainty 
$\delta X^{\rm NLO}$ to the NLO prediction $X^{\rm NLO}$ 
of a given observable $X$ is \cite{Epelbaum:2014efa,Binder:2016bi}
\begin{eqnarray}
\label{Error}
\delta X^{\rm NLO} (Q) &=& \max  \bigg( Q^3 \times \Big| X^{\rm
    LO}(Q) \Big|, \cr
  && Q \times \Big|
  X^{\rm LO}(Q) -   X^{\rm NLO}(Q) \Big|\bigg) .
\end{eqnarray}
We also note that the additional constraints specified
in Eq.~(8) of Ref.~\cite{Binder:2016bi} are imposed. 
In Refs.~\cite{Epelbaum:2019zqc,Epelbaum:2019wvf}, the expansion parameter $Q$
was estimated to be $Q=0.31$. This value was also used in nucleonic few-body  studies~\cite{Epelbaum:2019zqc,Maris:2020qne}. 
We adopt here also the value of $Q=0.4$ obtained in the Bayesian analysis,
cf. Sect.~\ref{Sec:Potentials2}. 
Using this ansatz to estimate the uncertainty, we obtain the results listed in Table~\ref{tab:ekm}
for the uncertainties due to the truncation. 

In Ref.~\cite{Melendez:2017phj}, it was found that the EKM uncertainty estimates correspond at NLO to the 68\% DoB interval for a specific 
choice of the prior. Here, we find that the values are of similar order as the Bayesian analysis. Choosing the same expansion coefficient 
as in our Bayesian analysis, the actual values are somewhat larger. 
Only for the standard choice $Q=0.31$, we find good agreement between the two
uncertainty estimates. 

\begin{table}[tbp]
    \begin{center}
    \renewcommand{\arraystretch}{1.4}
    \setlength{\tabcolsep}{0.4cm}        
    \vspace{1cm}
    \begin{tabular}{|r|rrrr|}
    \hline
         $Q$  & \nucl{3}{\Lambda}{H} & \nucl{4}{\Lambda}{He(0$^+$)} & \nucl{4}{\Lambda}{He(1$^+$) }  & \nucl{5}{\Lambda}{He} \\
         \hline 
         0.31  & 0.015 & 0.30  & 0.36  & 1.1   \\
         0.40  & 0.015 & 0.39  & 0.47  & 1.4  \\
         \hline
         0.31  & 0.005 & 0.06 & 0.07 & 0.64 \\
         0.40  & 0.008 & 0.13 & 0.09 & 0.83 \\
         \hline 
    \end{tabular}
    \caption{\label{tab:ekm} EKM uncertainty estimates in MeV at order NLO 
    using different $Y\!N$ (1st and 2nd line) and $N\!N$ (3rd and 4th line) orders 
    for \nucl{3}{\Lambda}{H}, \nucl{4}{\Lambda}{He}, \nucl{5}{\Lambda}{He} for two values of the expansion parameter $Q$.  }
    
    \end{center}
\end{table}

\bibliographystyle{unsrturl}

\bibliography{bibliography.bib}

\end{document}